\documentclass[10pt,onecolumn]{IEEEtran}

\input{epsf}
\usepackage{framed}
\usepackage{color}

\definecolor{bgrd}{rgb}{1,1,1}
\definecolor{grey}{rgb}{0.9,0.9,0.6}
\definecolor{gray}{rgb}{0.5,0.5,0.5}
\definecolor{dkr}{rgb}{0.6,0.2,0.2}
\definecolor{dkg}{rgb}{0,0.5,0}
\definecolor{dkb}{rgb}{0.0,0.1,0.7}
\definecolor{light-gray}{gray}{0.85}

\newcommand{\test}{\mbox{$\begin{array}{c}
\stackrel{\stackrel{\textstyle {\cal H}_1}{\textstyle >}}
{\stackrel{\textstyle <}{\textstyle {\cal H}_0}} \end{array}$}}

\renewcommand{\P}{\mathbb{P}}

\newcommand{\beq}{\begin{equation}}
\newcommand{\eeq}{\end{equation}}
\newcommand{\beqa}{\begin{eqnarray}}
\newcommand{\eeqa}{\end{eqnarray}}
\newcommand{\dfz}{\triangleq}
\newcommand{\deq}{\stackrel{d}{=}}

\newcommand{\bb}{{\mathbf{b}}}

\newcommand{\bu}{{\mathbf{u}}}
\newcommand{\bz}{{\mathbf{z}}}
\newcommand{\bg}{{\mathbf{g}}}

\newcommand{\by}{{\mathbf{y}}}

\newcommand{\bw}{{\mathbf{w}}}
\newcommand{\bv}{{\mathbf{v}}}

\newcommand{\bx}{{\mathbf{x}}}

\newcommand{\br}{{\mathbf{r}}}

\newcommand{\V}{\mathbb{V}}

\newcommand{\E}{\mathbb{E}}

\newcommand{\cE}{{\cal E}}

\newcommand{\cI}{{\cal I}}

\newcommand{\cN}{{\cal N}}

\newcommand{\cH}{{\cal H}}
\usepackage{cite}
\usepackage{multirow}
\usepackage{rotating}
\usepackage{hhline}
\usepackage{amssymb}
\usepackage{mathrsfs}
\usepackage{graphicx}
\usepackage{epstopdf}
\usepackage{amsmath}
\usepackage{empheq}

\usepackage{soul}
\usepackage[vlined,ruled]{algorithm2e}

\newcommand{\N}{{\cal N}}

\usepackage{xcolor, framed}

\newenvironment{myframedeq}[1][\linewidth]{\FrameSep=4pt\abovedisplayskip=0pt\belowdisplayskip=0pt
\framed\hsize=#1\leftskip=\dimexpr(\textwidth-#1)/2\relax}
{\endframed}

\begin{document}

\title{Detection under One-Bit Messaging \\ over Adaptive Networks} 

\author{Stefano~Marano and Ali H. Sayed, \IEEEmembership{Fellow, IEEE}
\thanks{S.~Marano is with DIEM, University of Salerno, via Giovanni Paolo~II 132, I-84084, Fisciano (SA), Italy (e-mail: marano@unisa.it). A.~H.~Sayed is with the Ecole Polytechnique Federale de Lausanne EPFL, School of Engineering, CH-1015 Lausanne,
Switzerland (e-mail: ali.sayed@epfl.ch).}
\thanks{The work of A.~H.~Sayed was also supported in part by NSF grant CCF-1524250.}
\thanks{Part of this work has been presented at EUSIPCO 2018~\cite{eusipco18}.}}

\maketitle

\begin{abstract}

This work studies the operation of multi-agent networks engaged in binary decision tasks, and derives performance expressions and performance operating curves under challenging conditions with some revealing insights. One of the main challenges in the analysis is that agents are only allowed to exchange one-bit messages, and the information at each agent therefore consists of both continuous and discrete components. Due to this mixed nature, the steady-state distribution of the state of each agent cannot be inferred from direct application of central limit arguments. Instead, the behavior of the continuous component is characterized in integral form by using a log-characteristic function, while the behavior  of the discrete component is characterized by means of  an asymmetric Bernoulli convolution.  By exploiting these results, the article derives reliable approximate performance expressions for the network nodes that match well with the simulated results for a wide range of system parameters. The results also reveal an important  interplay between continuous adaptation under constant step-size learning and the binary nature of the messages exchanged with neighbors. 

\end{abstract}


\section{Introduction}
\label{sec:intro}

\IEEEPARstart{T}{he} theory of adaptive decision systems lies at the intersection of the fields of decision theory~\cite{Lehmann-testing3,poorbook} and adaptive learning and control~\cite{Sayed2008adaptive,Astrom-Wittenmark}. 
In many instances, the qualification ``adaptive'' in adaptive decision systems refers to the ability of the system to select the best action based on the observed data~\cite{chernoff-sequential} as in cognitive radar~\cite{haykincognitiveradar}, active hypothesis testing~\cite{Javidi-AS}, or controlled sensing~\cite{Veeravalli-active-AC,FMM-Chernoff17}. The qualification ``adaptive'' can also refer to the ability of the decision system to track changes in the underlying state of nature and to monitor its drifts over time and deliver reliable decisions in real time. The theme of this article falls into this second type of decision systems.

The classical implementation of adaptive decision systems has often relied on the use of centralized (fusion) processing units. In more recent years, there has been a shift from centralized architectures to sensor network architectures~\cite{pradhan02,akyildiz-survey,chong&kumar,SPMAGAZINE-special,PreddKulkarniPoor-magazine06,chamberland,ChenTongVarshneyMag} where data are monitored/collected in a distributed fashion by a collection of individually simple devices but the processing continues to be centralized. Energy efficiency, robustness, and security issues become a challenge over such implementations. Besides, the presence of a single central unit makes the system vulnerable to failures and external attacks~\cite{akyildiz-survey,chong&kumar,appadwedulaJSAC,BlumSadlerSP08}. One approach to remedy these difficulties is the SENMA (sensor networks with mobile agents) paradigm in which several mobile central units travel across the network to query the remote nodes from close proximity~\cite{SENMA,tong:C-SENMA-LDPC,doasplet05,spawc06}.

A more prominent and flexible solution is to avoid the presence of central units. 
A fully-flat or fully-decentralized sensor network refers to a network in which all the information processing takes place at the nodes in a fully distributed fashion, and no data storage or processing at centralized devices is allowed. The evident mitigation of security and failure issues, and the added robustness, come at the expense of the need for local information processing capabilities at the nodes, which will now need to interact with each other, perform processing tasks with groups of nearby agents, and arrive at local decisions. 
Statistical signal processing over fully-decentralized networks or graphs has become an active area of research (e.g., see the overviews in~\cite{SayedNOW2014,SayedprocIEEE} and the many references therein). The theme of this work is to design and analyze an adaptive decision system over such networks.

\subsection{Related Work}

Inference problems over fully-flat sensor networks have received considerable attention in the last years in connection with estimation problems first 
and, more recently, in connection with detection/decision 
problems by employing either consensus~\cite{kar-moura-stsp,running-cons,asymptotic-rc,mouraetal2011,mouraetal2012,mouraetal2012_2,boyd-infocom} or diffusion~\cite{CattivelliSayedEstimation,ZhaoSayedLMSestimation,TuSayedConsensus,CattivelliSayedDetection,SayedSPmag,sayedDAoN,SayedprocIEEE,chen-sayed-IT1,chen-sayed-IT2,BracaetalIT,MattaSIPN16} strategies. 

Consensus solutions employ diminshing step-sizes to enhance the memory properties of the decision system, 
which leads to asymptotically optimal performance~\cite{kar-moura-stsp,running-cons,asymptotic-rc,mouraetal2011,mouraetal2012,mouraetal2012_2}.
Unfortunately, decaying step-size parameters limit the adaptation ability of the resulting network because learning comes to a halt as the step-size parameter approaches zero. Switching to constant step-size adaptation poses a challenge for consensus-based solutions because of an inherent asymmetry in the update equations of consensus implementations. This asymmetry has been studied in some detail and shown earlier in the works~\cite{SayedNOW2014,TuSayedConsensus} to be a source of instability when constant step-sizes are used for adaptation purposes. In other words, consensus strategies under constant step-sizes can be problematic for applications that necessitate continuous learning due to potential instability. This fact motivates us to focus on diffusion implementations since these strategies do not suffer from the aforementioned asymmetry and have been shown to deliver superior performance under both constant and decaying step-size learning scenarios~\cite{SayedNOW2014,TowficChenSayedIT2016}.
There have been a series of works that develop the theory of diffusion strategies with constant step-size for inference purposes and explore their capabilities for learning and adaptation in dynamic environments~\cite{CattivelliSayedEstimation,ZhaoSayedLMSestimation,TuSayedConsensus,CattivelliSayedDetection,SayedSPmag,sayedDAoN,SayedprocIEEE,chen-sayed-IT1,chen-sayed-IT2,BracaetalIT,MattaSIPN16}. 
For example, references~\cite{CattivelliSayedEstimation,ZhaoSayedLMSestimation,TuSayedConsensus} deal with estimation problems, and the latter also addresses a comparison between consensus and diffusion protocols. In~\cite{CattivelliSayedDetection} the adaptive diffusion scheme for detection is studied, 
and~\cite{SayedSPmag,sayedDAoN,SayedprocIEEE} present extensive overviews of these detection algorithms, as well as many access points to the related literature. The learning behavior of the network is investigated in~\cite{chen-sayed-IT1,chen-sayed-IT2}, while a large deviation analysis and the so-called exact asymptotic framework are the focus of~\cite{BracaetalIT,MattaSIPN16}.

\subsection{Contribution \& Preview of the Main Results}

The diffusion scheme considered in this paper employs a modified form of the adapt-then-combine (ATC) diffusion rule, which has some advantages
with respect to alternative schemes~\cite{SayedSPmag}. According to the ATC rule, each node updates (adapts) its state by incorporating the fresh information coming from new measurements, and then makes its current state available to its neighbors for the combination stage. In the combination stage each node weighs its state with those of its neighbors. In all the articles mentioned so far in this introduction it is assumed that the communication among nearby nodes is essentially \emph{unconstrained}. This means that the nodes can share their state with the neighbors with full precision.

In many practical scenarios, an unconstrained inter-node communication capability cannot be guaranteed, and the system designer is faced with the problem of revisiting the signal/information processing of the network in order to take into account this limitation. 
Thus, the basic consideration that motivates our work is that the nodes of the network cannot exchange their states as they are, because the communication links do not support messages with arbitrary precision. Taking this viewpoint to one extreme, we assume severe communication constraints which impose that only one bit can be reliably delivered, per link usage, over the inter-node links. Accordingly, in the combination stage of the ATC rule, the neighbors of node~$k$ cannot be informed about the value of the state of node~$k$, but they can only be informed about a one-bit quantized quantity. In the binary detection problem addressed here, this quantized quantity can be regarded as a local decision made by node~$k$ at the current time.

Of course, decentralized inference using quantized messages in networks equipped with a central unit (or having other classical structures, such as the tandem architecture or some variation thereof) has a long history, see e.g.,~\cite{Tsitsiklis88,Varshney:book,Luo-univdetect,Ma-Ma-Ma,VarshneySP12} and the references therein. Also, consensus implementations with quantized messages has been widely investigated~\cite{BiaoChenISIT16,ZhuChen2016,ZhuChen2016-corr,Basar-2016-AC,ZhuSohXie,Lietal-2011,Kar-Moura2010,Carlietal2010,AysalCoatesRabbat}. Apparently there are not similar studies on distributed strategies ensuring continuous learning and adaptation even under drifting or non-stationary conditions.

The one-bit diffusion messaging scheme addressed in this paper poses new challenges. The combination stage of the ATC scheme will now fuse discrete and continuous variables and the analysis of the steady-state distribution becomes more complex than that developed, e.g., in~\cite{BracaetalIT}. In contrast to the results of~\cite{BracaetalIT}, our analysis shows that (a version of) the central limit theorem (CLT) only applies in the special circumstance that the step size is small and the weight assigned to local decisions gathered by neighbor nodes is vanishing. 
In general cases, for arbitrary step sizes and combination weights, 
deriving the steady-state statistical distributions 
requires separate analysis of the continuous and discrete components. Neither of these can be, in general, approximated by a Gaussian distribution via some version of the CLT, and different analysis tools are required.

The main results of this work can be summarized as follows. By exploiting a key distributional structural property [see~(\ref{eq:key2})], the steady-state distribution of the continuous component [see~(\ref{eq:winf})] is obtained in an integral form that involves the log-characteristic function.
A series approximation for the continuous component distribution, particularly suitable for numerical analysis, is also provided. These results are collected in Theorem~1.

As to the discrete component, 
we show that this reduces to a combination of geometric series with random signs ---the so-called asymmetric Bernoulli convolution. The
Bernoulli convolution has been widely studied in the literature for its measure-theoretic implications and it is known that, aside from some special cases, its distribution does not reduce to simple forms. This notwithstanding, exploiting the fact that the node state is the sum of the discrete component with the continuous one, we derive simple approximations in the regime of highly reliable local decisions [$1-p_d,p_f \ll 1$,  see~(\ref{eq:pdpf})]. In principle, approximations of any degree can be developed, but the second-order approximation detailed in Sec.~\ref{sec:soap} gives accurate results even for moderately large values of $1-p_d$ and $p_f$.
The combination of several Bernoulli convolutions requires numerical convolutions and we develop careful approximations for the individual contributions so that these convolutions can be easily computed over discrete, low-cardinality, sets.

Exploiting the above results we finally get reliable approximate expressions for the steady-state distributions of the network nodes. 
Our analysis highlights the role of the system parameters on the these distributions, and the inherent system tradeoffs.
The examples of computer experiments (reported in Figs.~\ref{fig:gausH0}-\ref{fig:expoH1}) show that the shape of the steady-state distributions is by no means obvious. It is rewarding that the developed theory is able to closely follow those shapes for a wide range of values of the relevant system parameters. 
The main performance figures are the system-level detection and false alarm probabilities $P_d$ and $P_f$, which are straightforwardly related to the distributions of the network nodes. Expressing $P_d$ in function of $P_f$, the receiver operating characteristic (ROC) curve is obtained (Figs.~\ref{fig:rocg}-\ref{fig:roce}).

The analysis developed in this paper allows us to easily derive the decision performance of the system for a wide range of the parameters under the control of the system designer --- step size $\mu$ and combination weights $\{a_k\}$. 
A critical scenario is when the self-combination weights~$\{a_k\}$ are very large and $\mu$ is very small, because the developed numerical procedures can be time-consuming. Furthermore, for $a_k\to1$ and $\mu\to0$, both the continuous and the discrete components 
may become indeterminate and a joint analysis of these components is necessary. For this scenario we develop a tailored version of the CLT for triangular arrays and continuous parameters, which is the subject of Theorem~2.

From a practical perspective, our main results are a formula (shown just after Theorem~1) providing a simple numerical recipe for the steady-state distributions of the continuous component, and an approximation for the discrete component, which are simply combined to yield the final distributions of the agents at the steady-state. Using these results the system designer can tune the agents' thresholds in order to achieve a desired value of system-level $P_f$, and compute the corresponding $P_d$.

The remainder of this paper is organized as follows. Section~\ref{sec:ADD} introduces the classical adaptive diffusion scheme for detection. The one-bit-message version of these detection systems is designed in Sec.~\ref{sec:one-bit}, and the steady-state analysis is conducted in Sec.~\ref{sec:SSA}. 
Examples of applications of the developed theory and results of computer experiments are presented in Sec.~\ref{sec:compe}. Extensions of the proposed approach are briefly discussed in Sec.~\ref{sec:added}, while Sec.~\ref{sec:concl} concludes the paper with final remarks. Some technical material is postponed to Appendices~\ref{app:newappth1}-\ref{app:delta}.

\section{Adaptive Diffusion for Detection}
\label{sec:ADD}
We consider a multi-agent network consisting of $S$ nodes running an adaptive diffusion scheme to solve
a binary hypothesis test problem in which the state of nature is represented by $\cH_0$ or $\cH_1$.
Using the same notation from~\cite{BracaetalIT}, the update rule for the diffusion strategy is given by
\begin{subequations} \begin{align}
&\bv_k(n)=\by_k(n-1)+\mu [\bx_k(n)-\by_k(n-1)], \label{eq:old1}\\
&\by_k(n)= \sum_{\ell=1}^S a_{k \ell} \bv_{\ell}(n), \qquad n\ge 1\label{eq:old2}, 
\end{align} \end{subequations}
where $0 < \mu \ll 1$ 
is the step-size parameter, usually much smaller than one. Moreover, the symbol $\bx_k(n)$ denotes the data received by agent $k$ at time $n$, while $\by_k(n)$ represents a local state variable  that is updated regularly by the same agent through~(\ref{eq:old2}). This latter expression combines the intermediate values $\bv_{\ell}(n)$ from the neighbors of agent $k$ using the nonnegative convex combination weights $\{a_{k\ell}\}$. The weights are required to satisfy
\beq
a_{k\ell}\geq 0,\quad \sum_{\ell\in{\cal N}_k} a_{k\ell}=1,
\eeq
where ${\cal N}_k$ denotes the set of neighbors connected to agent~$k$, including $k$ itself. In the above notation, the scalar $a_{k\ell}$ denotes the weight by which information flowing from $\ell$ to~$k$ is scaled. If agents $k$ and $\ell$ are not neighbors, then $a_{k\ell}=0$. Expressions~(\ref{eq:old1})-(\ref{eq:old2}) can be grouped together across all agents in vector form, say, as:
\begin{subequations} \begin{align}
&\bv_n = (1-\mu) \by_{n-1}+ \mu \bx_n, \label{eq:bvold}\\
&\by_n = A \bv_n, \qquad n \ge 1, \label{eq:byold}
\end{align} \end{subequations}
where the combination matrix $A=[a_{k \ell}]$ is $S \times S$, while 
$\by_n=\mbox{\rm col}\{\by_1(n), \by_2(n),\ldots,$ $\by_S(n)\}$. Similarly for $\bv_n$ and~$\bx_n$. Iterating~(\ref{eq:bvold})-(\ref{eq:byold}) gives
\beq
\by_n = (1-\mu)^n  A^n  \by_{0} + \mu \sum_{i=0}^{n-1} (1-\mu)^i A^{i+1} \bx_{n-i} . 
\label{eq:recold}
\eeq

In this work we make the assumption that the incoming data $\bx_{k}(n)$ is a statistic computed from
some observed variable, say, $\br_{k}(n)$, namely, $\bx_{k}(n)$ is a prescribed function of $\br_{k}(n)$.
Under both hypotheses $\cH_0$ and $\cH_1$, the observations $\br_k(n)$ are i.i.d.\ (independent and identically distributed) across all sensors $k=1,\dots,S$, and over time.
It follows that the same i.i.d.\ property holds for $\{\bx_k(n)\}$.
We further assume that each $\bx_k(n)$ is an absolutely continuous random variable having a probability density function (PDF) with respect to the Lebesgue measure~\cite{billingsley-book2}, under both hypotheses. This assumption is mainly because the case of continuous random variables is the most interesting in the presence of data quantization. 

We refer to $\br_k(n)$ as the local observation, and to $\bx_k(n)$ as the marginal decision statistic, where the adjective ``marginal'' is meant to indicate that $\bx_k(n)$ is based on the single sample $\br_k(n)$.
The variable $\by_k(n)$ is referred to as the \emph{state} of the node. The detection problem consists of comparing the state $\by_k(n)$ against a threshold level, say~$\gamma \in \Re$, and deciding on the state of nature $\cH_0$ or $\cH_1$, namely,
\beq
\by_k(n) \test \gamma.
\label{eq:test}
\eeq

While our formulation is general enough to address different types of  marginal statistics, special attention will be given to the case in which $\bx_{k}(n)$ is selected as the log-likelihood ratio of $\br_k(n)$:
\beq
\bx_{k}(n)=\log \frac{f_{\br,1}(\br_k(n))}{f_{\br,0}(\br_k(n))}, 
\label{eq:logli}
\eeq
where $f_{\br,h}(\br_k(n))$ is the PDF of $\br_k(n)$ under $\cH_h$, $h=0,1$.

\subsection{Some Technical Conditions}
We introduce the following technical conditions. First, we let $\E_h$ denote
the expectation under hypothesis $\cH_h$, $h=0,1$, and assume that
$-\infty<\E_0 \bx < \E_1 \bx < \infty$. 
The assumption $\E_0 \bx < \E_1 \bx $ is automatically verified when the marginal statistic is the log-likelihood ratio~(\ref{eq:logli}) because, in that case, the quantities $-\E_0 \bx$ and  $\E_1 \bx$ are two Kullback-Leibler divergences and, therefore, they are strictly positive for distinct hypotheses~\cite{CT2}.
We also assume that the variance $\V_h \bx$ exists and is finite for $h=0,1$.
Note that, for simplicity, we are using the short-hand notation $\bx$ instead of $\bx_k(n)$. 

Second, we let $\P_h$ denote the probability operator under hypothesis $\cH_h$, $h=0,1$, 
and assume, for all agents $\ell=1,2,\ldots,S$, that
\beq
0 < p_f   \dfz  \P_0(\bx \ge \gamma_{\rm loc}) < \P_1(\bx \ge \gamma_{\rm loc}) \dfz p_d < 1,
\label{eq:pdpf}
\eeq 
where $\gamma_{\rm loc}$ is a \emph{local} threshold level, and where $p_d$ and $p_f$ represent the \emph{marginal} detection and false alarm probabilities, namely, the probabilities that would be obtained 
by making decisions based on the marginal statistic and the local threshold:
$\bx_{\ell}(n) \ge \gamma_{\rm loc}$ $\Rightarrow$ decide locally in favor of $\cH_1$, otherwise decide for $\cH_0$. Note that $\gamma_{\rm loc}$ is  the same for all sensors, which is justified by the assumption of i.i.d.\ observations under both hypotheses. As a consequence, all sensors have the same \emph{marginal} performance $p_d$ and $p_f$.
The assumption $p_f,p_d \neq 0,1$, in~(\ref{eq:pdpf}) rules out trivialities. The condition $p_f<p_d$ is known as unbiasedness of the marginal decisions~\cite{Lehmann-testing3}.

\section{One-bit Diffusion Messaging} 
\label{sec:one-bit}

The classical diffusion rule is described by equations~(\ref{eq:old1})-(\ref{eq:old2}), and its detection properties are studied in detail in~\cite{BracaetalIT,MattaSIPN16}. In the system described by~(\ref{eq:old1})-(\ref{eq:old2}), the data exchanged among the nodes are 
uncompressed and non-quantized. However, in most sensor network scenarios, a 
more realistic assumption is that the messages exchanged among the nodes are quantized. 
Accordingly, we will consider an update rule in which the information sent at time~$n$ by node $\ell$ to its neighbors is the marginal statistic $\bx_{\ell}(n)$ \emph{quantized to one bit}, as follows:
\beq
\widetilde \bx_{\ell}(n) \dfz \left \{ \begin{array}{ll} \E_1 \bx, & \textnormal{ if } \bx_{\ell}(n) \ge \gamma_{\rm loc}, \\ \E_0 \bx, & \textnormal{ if } \bx_{\ell}(n) < \gamma_{\rm loc}, \end{array} \right.
\label{eq:q}
\eeq
where $\gamma_{\rm loc}$ is the local threshold level from~(\ref{eq:pdpf}).
Thus, if the state of nature is $\cH_1$, node $\ell$ sends to its neighbors the message $\E_1 \bx$ with probability $p_d$, and the message $\E_0 \bx$ with probability $(1-p_d)$. Similarly, under $\cH_0$, $\E_1 \bx$ is sent with probability $p_f$, and $\E_0 \bx$ with probability $(1-p_f)$. Needless to say, given that the nodes are aware of the detection problem they are faced with, there is no need to deliver the actual values $\E_{0,1} \bx$, but simply a binary flag.
We can interpret $\widetilde \bx_{\ell}(n)$ as representative of the \emph{marginal} decision about the hypothesis, made by node~$\ell$ by exploiting only its current observation $\br_{\ell}(n)$. 

When the nodes compute the log-likelihood ratio of the 
observations using~(\ref{eq:logli}), we can set $\gamma_{\rm loc}=0$ in~(\ref{eq:pdpf}) and~(\ref{eq:q}).
Comparing the log-likelihood ratio of $\br_{\ell}(n)$ to zero corresponds to the optimal ML (Maximum Likelihood) \emph{marginal} decision about the underlying hypothesis~\cite{kaydetection}.

Formally, instead of~(\ref{eq:old1})-(\ref{eq:old2}), we consider the following update rule with the one-bit messages: 
\begin{myframedeq}
\begin{subequations}
  \begin{align}
&\bv_k(n)=\by_k(n-1)+\mu [\bx_k(n)-\by_k(n-1)], \label{eq:new1}\\
&\by_k(n)= a_{kk} \bv_k(n)  + \sum_{\ell\neq k} a_{k \ell}\, \widetilde \bx_{\ell}(n), \qquad n \ge 1. \label{eq:new2}
  \end{align}
  \end{subequations}
\end{myframedeq}

Note that in the scheme of~(\ref{eq:new1})-(\ref{eq:new2}) sensor $\ell$ sends to its neighbors the quantized version $\widetilde \bx_{\ell}(n)$ of the marginal decision statistic. 
One alternative would be a system in which the sensor sends to its neighbors the quantized version $\widetilde \bv_{\ell}(n)$ of its state. In this case, however, it would not be possible to derive a simple analytical relationship between $\by_k(n)$ and $\by_k(n-1)$ that yields an explicit expression for the state $\by_k(n)$ [as in~(\ref{eq:rec}) below], and the analytical tractability would be compromised.
More importantly, the scheme~(\ref{eq:new1})-(\ref{eq:new2}) ensures improved \emph{adaptation} performance properties. 
Indeed, by using~(\ref{eq:new1})-(\ref{eq:new2}), the quantity $\bx_{\ell}(n)$ available at node $\ell$ will be used [through its quantized version $ \widetilde \bx_{\ell}(n)$] by the set of neighbors of node~$\ell$ (even though it is never made available to non-neighboring nodes). Changes in the state of nature are immediately reflected in the value of $\bx_{\ell}(n)$, while, with $\mu\ll1$, the state $\bv_{\ell}(n)$ of the node incorporates these changes only slowly, as shown in~(\ref{eq:old1}) and~(\ref{eq:new1}).
In dynamic environments, where the state of nature changes with time, this means that the system described by~(\ref{eq:new1})-(\ref{eq:new2}) will be able to react more rapidly to these changes, especially when the self-combination coefficient $a_{kk}$ is small.

\begin{figure}
\centering 
\includegraphics[width =190pt]{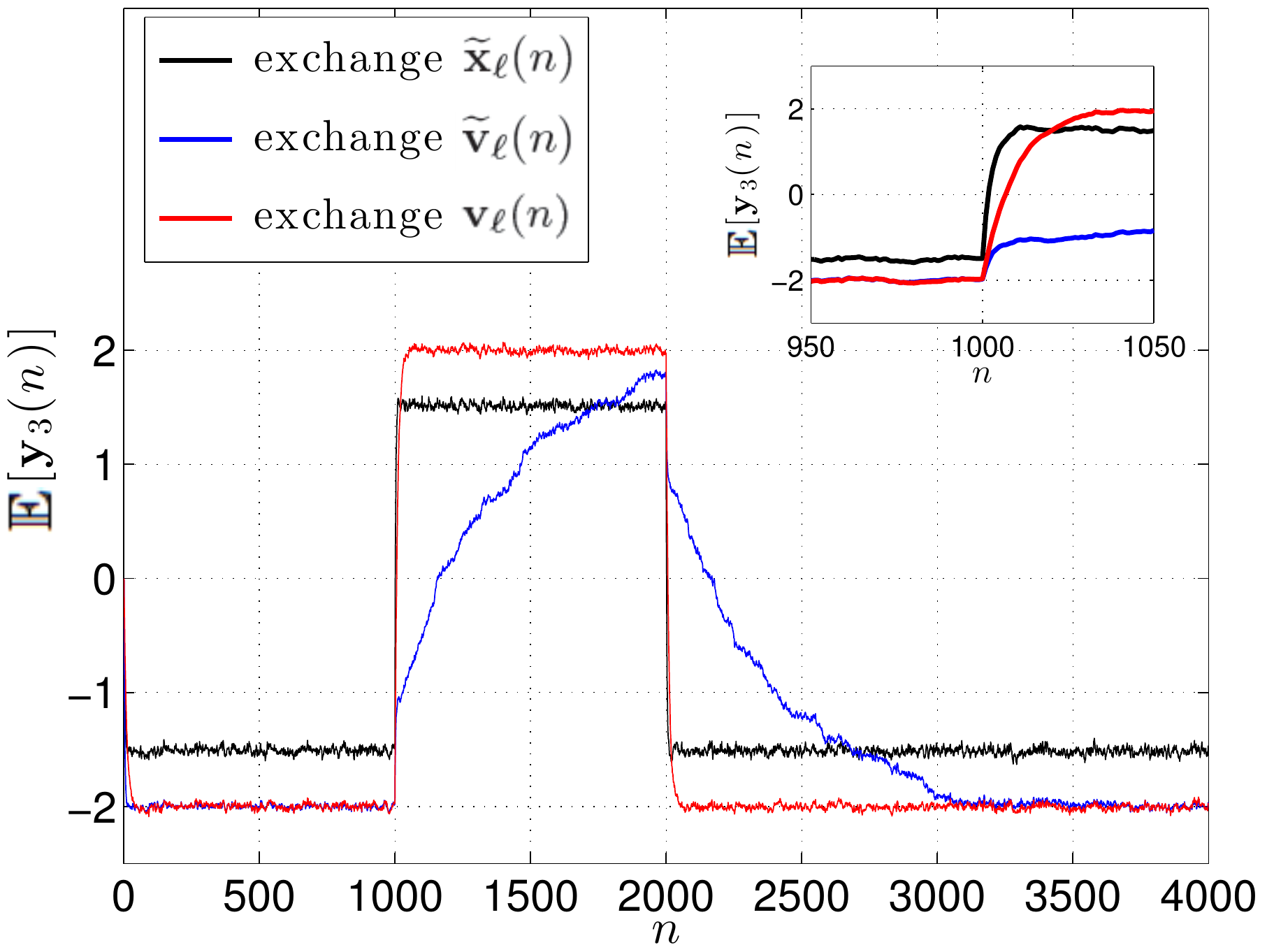}
 \caption{Example of the adaptive properties of the diffusion scheme~(\ref{eq:new1})-(\ref{eq:new2}) (curve in black), as compared to a system that exchanges the quantized state $\widetilde \bv_\ell(n)$ (blue), and to a system that exchanges the unquantized state $\bv_\ell(n)$ (red). The state of nature is $\cH_0$ from $n=1$ to $n=1000$, then switches to $\cH_1$ up to $n=2000$, and finally switches back to $\cH_0$.
The curves show the evolution of the expected value $\E[\by_k(n)]$ for node $k=3$ for the network shown in Fig.~\ref{fig:net}, when the $\bx$'s are Gaussian random variables distributed as detailed in the example of Sec.~\ref{sec:gau}, with $\rho=2$. The combination matrix is as in~(\ref{eq:lap}), with $a_k=0.75$, and the step size is $\mu=0.1$. The expectations are computed by averaging 100 Monte Carlo runs. The inset zooms on the region around $n=1000$.}
      \label{fig:open}
\end{figure}

\begin{figure}
\centering 
\includegraphics[width =150pt]{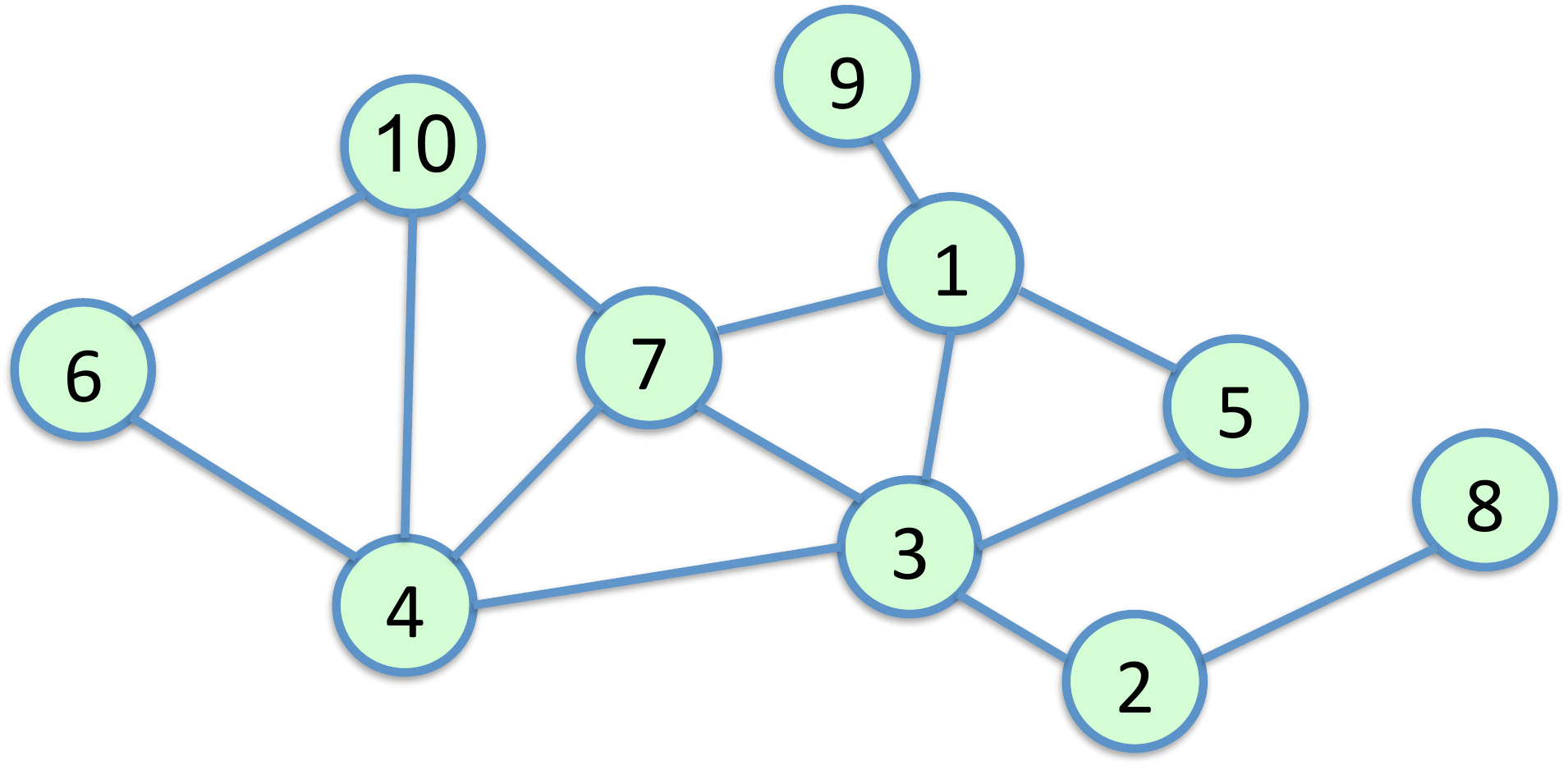}
 \caption{Topology of the network used in the computer experiments.}
      \label{fig:net}
 \end{figure}

Figure~\ref{fig:open} illustrates the improved adaptive properties of the scheme~(\ref{eq:new1})-(\ref{eq:new2}) by showing the expected value $\E[\by_k(n)]$ for node $k=3$ of the network shown in Fig.~\ref{fig:net}, when the $\bx$'s are Gaussian random variables distributed as detailed in the example of Sec.~\ref{sec:gau}. The state of nature is initially $\cH_0$, then switches to $\cH_1$ at $n=1000$, and finally switches down to $\cH_0$ at $n=2000$.
The solid curve in black represents $\E[\by_k(n)]$ for the system defined by recursion~(\ref{eq:new1})-(\ref{eq:new2}), where nodes exchange the quantized version $\widetilde \bx_\ell(n)$ of their marginal statistic, while the curve in blue refers to a one-bit message scheme in which the nodes exchange the quantized version $\widetilde \bv_\ell(n)$ of their state. It is evident that this latter system does not react promptly to changes in the underlying state of nature and therefore is less suitable to operate in dynamic environments, with respect to the system defined by~(\ref{eq:new1})-(\ref{eq:new2}).

For comparison purposes, Fig.~\ref{fig:open} also shows the expected value $\E[\by_k(n)]$ of a diffusion scheme in which no restrictions are imposed on the messages and therefore the nodes are allowed to exchange the \emph{unquantized} state $\bv_\ell(n)$. This is the diffusion system~(\ref{eq:old1})-(\ref{eq:old2}) studied in~\cite{BracaetalIT,MattaSIPN16}. The inset of Fig.~\ref{fig:open} 
makes it evident that the scheme of~(\ref{eq:new1})-(\ref{eq:new2}) exhibits faster reaction even in comparison to the message-unconstrained diffusion scheme~(\ref{eq:old1})-(\ref{eq:old2}). 
Therefore, investigating the steady-state detection performance of the system defined by~(\ref{eq:new1})-(\ref{eq:new2}) is of great importance, and is the main theme of this work.

\subsection{Explicit Form of the State $\by_k(n)$}

Let us introduce the coefficients
\beq
c_{k \ell}= \left \{ \begin{array}{ll} a_{k \ell}, & k \neq \ell, \\ 0, & k=\ell, \end{array} \right.
\label{eq:c}
\eeq
and, for notational simplicity, let us write $a_k$ in place of $a_{kk}$.
The $\{c_{k\ell}\}$ correspond to the off-diagonal elements of $A$. By iterating the expressions in~(\ref{eq:new1})-(\ref{eq:new2}), we arrive at
\beqa
\by_k(n)&=& [(1-\mu)a_k]^n \by_k(0) \nonumber \\
&& + \; \; \mu \, a_k \sum_{i=0}^{n-1}  (1-\mu)^ia_k^i \bx_k(n-i) \nonumber \\
&& + \; \; \sum_{i=0}^{n-1}  \sum_{\ell=1}^S (1-\mu)^i a_k^i c_{k \ell} \, \widetilde \bx_{\ell}(n-i).
\label{eq:rec}
\eeqa
For notational convenience, we also introduce the scalar:
\beq
\eta_k \dfz (1-\mu)a_k, \qquad \eta_k \in (0,1),
\label{eq:eta}
\eeq
which combines the step size $\mu$ with the self-combination coefficient $a_k$. 
It is useful to bear in mind that $\eta_k$ is defined as a combination of $\mu$ and $a_k$, although our notation does not emphasize that.

Consider the quantity $\by_k(n)$ in~(\ref{eq:rec}). After a change of variable $i\leftarrow i+1$ we have:
\begin{align}
\by_k(n)&= \underbrace{\eta_k^n \, \by_k(0)}_{\textnormal{transient}}
+ \underbrace{\mu \, a_k \sum_{i=1}^{n}  \eta_k^{i-1} \, \bx_k(n-i+1)}_{\dfz \bu_k(n)} \nonumber \\
& + \underbrace{\sum_{i=1}^{n}  \sum_{\ell=1}^S \eta_k^{i-1} \, c_{k \ell} \, \widetilde \bx_{\ell}(n-i+1)}_{\dfz \bz_k(n)}.
\label{eq:rec2}
\end{align}
For $n\rightarrow \infty$, the transient part converges exponentially to zero with probability one so that, by Slutsky 
theorem~\cite[Th.\ 11.2.11]{Lehmann-testing3}, 
$\lim_{n\rightarrow \infty} \by_k(n)$ converges in distribution to $\lim_{n\rightarrow \infty} [\bu_k(n) + \bz_k(n)]$. 
In the following, we will investigate the steady-state properties of the system described by Eqs.~(\ref{eq:new1})-(\ref{eq:new2}) and, accordingly, 
we can set without loss in generality, 
\beq
\by_k(0)=0,
\label{eq:yk0}
\eeq
so that the transient part $\eta_k^n \, \by_k(0)$ of~(\ref{eq:rec2}) is eliminated. Henceforth, if not stated explicitly otherwise, condition~(\ref{eq:yk0}) is always assumed. 

In the next section we analyze separately the two components $\bu_k(n)$ (referred to as the \emph{continuous} component) and $\bz_k(n)$ (the \emph{discrete} component), in the asymptotic regime of large number of iterations. This allows us to provide a suitable approximation for the statistical distribution of $\lim_{n\rightarrow \infty} \by_k(n)$, which is our goal.

\section{Steady-State Analysis}
\label{sec:SSA}

\subsection{Continuous Component $\bu_k(n)$}
\label{sec:u}

Consider the quantity $\bu_k(n)$ defined in (\ref{eq:rec2}). Since the random variables $\{\bx_k(n)\}$ are i.i.d., the following equality in distribution holds:
\beq
\lim_{n\to \infty}\sum_{i=1}^{n}  \eta_k^{i-1} \, \bx_k(n-i+1) \deq \sum_{i=1}^{\infty}  \eta_k^{i-1} \, \bx_k(i) \dfz \bw_k^\star ,
\label{eq:winf}
\eeq
and, therefore, 
\beq
\bu_k(\infty) \dfz \lim_{n \rightarrow \infty} \bu_k(n) \deq a_k \, \mu \, \bw_k^\star.
\label{eq:uinf}
\eeq
Let us introduce a new random variable $\bx_k(0)$, which is an independent copy of the random variable $\bx_k(n)$.
From definition~(\ref{eq:winf}) the following structural property of $\bw_k^\star$ immediately follows:
\beq
\eta_k \, \bw_k^\star + \bx_k(0) \deq \bw_k^\star.
\label{eq:key2}
\eeq
The forthcoming Theorem~1 exploits~(\ref{eq:key2}) to derive the distribution of the random variable $\bw_k^\star$ and therefore the distribution of the steady-state variable $\bu_k(\infty)$ defined in~(\ref{eq:uinf}), whose mean and variance are easily computed from~(\ref{eq:uinf}) and~(\ref{eq:key2}):
\beq
\E_h[\bu_k(\infty)] = \frac{a_k \mu}{1-\eta_k} \E_h \bx , \, \V_h[\bu_k(\infty)] = \frac{a_k^2\mu^2}{1-\eta_k^2} \V_h \bx .
\label{eq:mev}
\eeq

Let $j=\sqrt{-1}$ be the imaginary unit and, for $t \in \Re$, let 
\beqa
\Phi_{\bx,h}(t) &\dfz& \log \E_h  e^{j \,t\, \bx },\\
\Phi_{\bw,h}(t) &\dfz& \log \E_h e^{j \,t\, \bw_k^\star }, 
\label{eq:logcf}
\eeqa
be the log-characteristic functions~\cite{billingsley-book2} of $\bx_k(n)$ and $\bw_k^\star$, respectively, under hypothesis $\cH_h$, $h=0,1$.
Also, denote by $F_{\bu,h}(u)$, $u \in \Re$, and $\Phi_{\bu,h}(t)=\Phi_{\bw,h}(a_k  \mu \,  t)$ the cumulative distribution function (CDF) and the log-characteristic function of the steady-state continuous component $\bu_k(\infty)$, respectively, under $\cH_h$.
Denote by $\lfloor z \rfloor$ the largest integer $ \le z$, and by ${\rm Im}\{z\}$ the imaginary part of $z$.
Finally, to simplify the notation, let
\begin{align}
\Omega_n(u,\delta)= \frac{{\rm Im} \left \{
\exp \left [\Phi_{\bw,h}\left ( \frac{2n+1}{2}
\delta\right )  - j  \frac{u}{\mu a_k} \frac{2n+1}{2}
\delta \right ]  \right \}}{2n+1} . \label{eq:notation}
\end{align}

\vspace*{5pt}
\noindent \textbf{Theorem 1 (Distribution of $\bu_k(\infty)$):} 
\emph{The continuous component~$\bu_k(n)$ converges in distribution for $n\rightarrow \infty$, and the CDF $F_{\bu,h}(u)$ of its limit $\bu_k(\infty)$ can be characterized  as follows.}
\emph{Suppose that $F_{\bu,h}(u)$ admits a density. Suppose also that $\Phi_{\bx,h}(t)$ can be expanded in a power series with radius of convergence $0<\tau_{\bx,h}\le \infty$, namely:
\beq
\Phi_{\bx,h}(t) = \sum_{n=1}^\infty \varphi_{n,h} \, t^n, \qquad |t| < \tau_{\bx,h}.
\label{eq:serieshyp}
\eeq
Then we have the following results. \\
i) The log-characteristic function $\Phi_{\bw,h}(t)$ can be uniquely expanded in a power series with radius of convergence $\tau_{\bx,h}$:
\beq
\Phi_{\bw,h}(t)=\sum_{n=1}^{\infty} \frac{\varphi_{n,h}}{1-\eta_k^n}  \, t^n, \qquad |t| < \tau_{\bx,h}.
\label{eq:Phiw}
\eeq
ii) If $\tau_{\bx,h}=\infty$, we have the representation:
\beq
F_{\bu,h}(u)= \frac 1 2 - \frac {1}{2 \pi j} \int_{-\infty}^\infty \hspace*{-7pt} \exp\left \{ \sum_{n=1}^{\infty} \varphi_{n,h} \frac{(\mu \, a_k t)^n}{1-\eta_k^n}  - j ut \right \} \frac{dt}{t},
\label{eq:newth}
\eeq
and the series appearing in~(\ref{eq:newth}) is (absolutely and uniformly) convergent for $t \in \Re$.}\\
\emph{iii) For $0<\tau_{\bx,h} \le \infty$, and $\delta>0$:} 
{\small 
\begin{align}
&\hspace*{-10pt} F_{\bu,h}(u) = \frac 1 2 - \frac {2}{\pi}
\sum_{n=0}^{\lfloor \frac{\tau_{\bx,h}}{\eta_k  \delta}-1 \rfloor}  \frac{1}{2n+1} \, \, {\rm Im} \Bigg \{ \exp \Bigg [-j \frac{u \, \delta}{\mu a_k}\frac{2n+1}{2} \nonumber \\
& \hspace*{-10pt} +  \Phi_{\bx,h} \hspace*{-2pt} \left (\frac{2n+1}{2}\delta \hspace*{-2pt} \right ) \hspace*{-2pt}
+ \hspace*{-3pt} \sum_{m=1}^{\infty}  \frac{\eta_k^m \, \varphi_{m,h}}{1-\eta_k^m} \left ( \frac{2n+1}{2}
\delta\right )^m 
\Bigg ] \hspace*{-2pt} \Bigg  \} + \Delta_{\bu}(u,\delta).
\label{eq:appnew}
\end{align}}%
\emph{Fix, $\epsilon^\prime>0$. If $\delta$ at the right-hand side of~(\ref{eq:appnew}) verifies 
\begin{align} \hspace*{-10pt}\left \{ \begin{array}{l}
\frac 2 \pi \Bigg |  \displaystyle{\sum_{n=0}^\infty} \Omega_n(u,\delta) 
-  \displaystyle{\sum_{n=0}^{\lfloor \frac{\tau_{\bx,h}}{\eta_k  \delta}-1 \rfloor} }\Omega_n(u,\delta)  \Bigg | \le \frac{ \epsilon^\prime}{2},
\vspace*{5pt}\\ 
  \hspace*{-2pt} \max \left \{ F_{\bu,h}  \left (u -  \frac{2 \pi a_k \mu}{\delta}  \right)   \hspace*{-2pt}, 
 1 \hspace*{-3pt}- \hspace*{-3pt} F_{\bu,h} \left(u + \frac{2 \pi a_k \mu}{\delta}  \right )
 \right \} \le \frac{ \epsilon^\prime}{2}  ,
\end{array}
\right .
\label{eq:maxnew}
\end{align}
where $\Omega_n(u,\delta)$ is defined in~(\ref{eq:notation}), then $|\Delta_{\bu}(u,\delta)| \le \epsilon^\prime$.}

\vspace*{3pt}
\noindent {\bf Proof:} See Appendix~\ref{app:newappth1}.\hfill$\square$  

\vspace*{3pt} 
The assumed existence of the density in Theorem~1 is known as absolute continuity of $F_{\bu,h}(u)$~\cite{billingsley-book2}, which holds for most distributions of practical interest. The series at the right-hand side of~(\ref{eq:serieshyp}) is well defined provided that $\Phi_{\bx,h}(t)$ is infinitely differentiable at the origin --- these derivatives are related to the cumulants of the random variable $\bx_k(n)$~\cite{papoulis}. Equality~(\ref{eq:serieshyp}) holds if $\Phi_{\bx,h}(t)$ is analytic in $|t| < \tau_{\bx,h}$~\cite{Zorichbook}.

Expression~(\ref{eq:appnew}) is used in practice by neglecting the term $\Delta_{\bu}(u,\delta)$ and truncating the series over $m$, yielding the approximation
\begin{align}
&\hspace*{-10pt} F_{\bu,h}(u) \approx \frac 1 2 - \frac {2}{\pi}
\sum_{n=0}^{\bar n}  \frac{1}{2n+1} \, \, {\rm Im} \Bigg \{ \exp \Bigg [-j \frac{u \, \delta}{\mu a_k}\frac{2n+1}{2} \nonumber \\
& \hspace*{-10pt} +  \Phi_{\bx,h} \hspace*{-2pt} \left (\frac{2n+1}{2}\delta \hspace*{-2pt} \right ) \hspace*{-2pt}
+ \hspace*{-3pt} \sum_{m=1}^{\bar m}  \frac{\eta_k^m \, \varphi_{m,h}}{1-\eta_k^m} \left ( \frac{2n+1}{2}
\delta\right )^m 
\Bigg ] \hspace*{-2pt} \Bigg  \} ,
\label{eq:appnew2}
\end{align}
for some $\bar n$, $\bar m$, and $\delta$.

To control the error, the first condition in~(\ref{eq:maxnew}) consists in truncating the series $\sum_{n=0}^\infty \Omega_n(u,\delta)$ to a sufficiently large integer $\bar n$.
In the proof of Theorem~1 it is shown that $\bar n$ and~$\delta$ must verify $(\bar n +\frac 1 2 )< \frac{\tau_{\bx,h}}{\eta_k  \delta}$. The value $\bar n=\lfloor \frac{\tau_{\bx,h}}{\eta_k  \delta}-1 \rfloor$ appearing in~(\ref{eq:appnew}) is accordingly chosen.
The second condition in~(\ref{eq:maxnew}) can be enforced by exploiting some estimate of the tails of $F_{\bu,h}(u)$, for which many methods are available~\cite{davies}. An example is given in Sec.~\ref{sec:compe}, see Appendix~\ref{app:delta}.

Since~(\ref{eq:appnew}) is not in simple analytical form, it is not immediate to understand how $F_{\bu,h}(u)$ depends on the system parameters, without resorting to numerical investigations as we shall do in Sec.~\ref{sec:compe}. However, from~(\ref{eq:mev}), the dispersion index can be computed:
\beq
\frac{\sqrt{\V_h[\bu_k(\infty)]}}{|\E_h[\bu_k(\infty)]|}=\frac{\sqrt{\V_h \bx}}{{|\E_h \bx|}} \sqrt{\frac{1-\eta_k}{1+\eta_k}},
\label{eq:di}
\eeq 
which reveals that the random variable $\bu_k(\infty)$ becomes more concentrated as~$\eta_k$ grows. Note that the second factor at the right-hand side of~(\ref{eq:di}) is not larger than unity, and depends on the parameters $a_k$ and $\mu$ only combined into~$\eta_k$.

One important remark is in order. In the special case that $\bx$ is a Gaussian random variable with mean $\E_h \bx$  and variance $\V_h \bx$, it is easily found that
$\Phi_{\bx,h}(t)= j \, t\, \E_h \bx - \frac 1 2 t^2 \, \V_h \bx$, see e.g.~\cite{papoulis}. From~(\ref{eq:Phiw}) one immediately gets
\beq
\Phi_{\bu,h}(t)= j \, t\,  \frac{a_k \mu}{1-\eta_k} \E_h \bx - \frac 1 2 \, t^2 \, \frac{a_k^2 \mu^2}{1-\eta_k^2} \V_h \bx , \quad t \in \Re,
\eeq
revealing that $\bu_k(\infty)$ is Gaussian with mean $ \frac{a_k \mu}{1-\eta_k} \E_h\bx $ and variance $\frac{a_k^2 \mu^2}{1-\eta_k^2} \V_h \bx$, see~(\ref{eq:mev}).
This appears to be an obvious result because, for all $n$, $\bu_k(n)$ is a linear transformation of the variables $\{\bx_k(n)\}$, and linear transformations preserve Gaussianity, see e.g.~\cite{papoulis}. 
For general (non-Gaussian) $\bx$, one might wonder if some form of CLT can be applied to the series appearing in~(\ref{eq:winf}), to infer that the distribution of $\bu_k(\infty)$ is Gaussian. This is not the case: One usual assumption for the CLT is that the sum of the variances of the summands diverges~\cite[Eq.~(8-123)]{papoulis}, whereas
 $\sum_{i=1}^n\eta_k^{2i-2} \V_h\bx$ converges to a finite value for $n\to \infty$.

\subsection{Discrete Component $\bz_k(n)$}
\label{sec:z}

We now derive an approximate distribution for the steady-state component $\lim_{n\rightarrow\infty} \bz_k(n)$, where $\bz_k(n)$ is defined in~(\ref{eq:rec2}). The approximation is valid for large $p_d$ and $(1-p_f)$ [see~(\ref{eq:pdpf}) for the definitions of these quantities], namely in the regime where the \emph{marginal} decisions of the nodes are reliable enough.

Let us start by an obvious equality in distribution:
\begin{align}
\bz_k(\infty) &\dfz \lim_{n\rightarrow \infty} \bz_k(n) = \sum_{\ell=1}^S c_{k \ell} \sum_{i=1}^{\infty}   \eta_k^{i-1}  \, \widetilde \bx_{\ell}(n-i+1) 
\label{eq:zinf} \\
& \deq \sum_{\ell=1}^S c_{k \ell} \sum_{i=1}^{\infty}  \eta_k^{i-1} \, \widetilde \bx_{\ell}(i) \; \dfz \; \bz_k^\star . \label{eq:begin} 
\end{align}
We introduce the normalized binary variables 
\beq
\bb_\ell(i) \dfz \frac{2\,  \widetilde \bx_{\ell}(i) - (\E_1 \bx+ \E_0 \bx)}{\E_1 \bx- \E_0 \bx}
\label{eq:bb}
\eeq
whose alphabet is $\{-1,+1\}$. When $\bx_k(n)$ is the log-likelihood ratio, as in~(\ref{eq:logli}), then $\bb_\ell(i)$ is simply the result of the signum function applied to $\widetilde \bx_{\ell}(i)$, because $\E_1 \bx>0$ and $\E_0 \bx<0$.
Using~(\ref{eq:bb}), the quantity $\bz_k^\star$ in~(\ref{eq:begin}) can be rewritten, after straightforward algebra, as
\begin{align}
\bz_k^\star =\sum_{\ell=1}^S  \frac{c_{k \ell}}{1-\eta_k} \small{\frac{(\E_1 \bx +\E_0 \bx) + (\E_1 \bx -\E_0 \bx)  
\,\bz_{k\ell}}{2}},
\label{eq:begin2}
\end{align}
where 
\beq
\bz_{k\ell} \dfz (1-\eta_k) \sum_{i=1}^{\infty} \eta_k^{i-1} \bb_\ell(i).  
\label{eq:rs}
\eeq
From~(\ref{eq:q}), note that under $\cH_1$ we have  $\P_1(\bb_\ell(i)=1)=p_d$, while, under $\cH_0$, $\P_0(\bb_\ell(i)=1)=p_f$.

Let us summarize some known properties of the series~(\ref{eq:rs}). Suppose that $\cH_1$ is in force.
If we had $p_d=1/2$, then expression~(\ref{eq:rs}) would represent a geometric series with equally likely random signs. This is known as the \emph{Bernoulli convolution}, and attracted considerable interest since the pioneering works by Erd{\"o}s~\cite{erdos39,erdos40}. 
It is known that the Bernoulli convolution is absolutely continuous with a finite-energy density for almost every $\eta_k \in (1/2,1)$, and is purely singular for $\eta_k \in (0,1/2)$, being in that case supported on a Cantor set of zero Lebesgue measure. It is also easily verified that for $\eta_k=1/2$ the random variable $\bz_{k\ell}$ is uniform in $(-1,1)$. We refer to~\cite{Peres-Solomyak-96} and the references therein for details.

Likewise, the \emph{asymmetric} Bernoulli convolution with $p_d \neq 1/2$, which is of interest to us, has been extensively studied. For $p_d \in (1/2,2/3)$, it is known that $\bz_{k\ell}$ is absolutely continuous for almost all $\eta_k > 2^{-H_b(p_d)}$, and singular for $\eta_k < 2^{-H_b(p_d)}$~\cite{Peres-Solomyak-98}, where $H_b(p)\dfz-p\log_2(p)-(1-p)\log_2(1-p)$ is the binary entropy function~\cite{CT2}. 
Obviously, the same considerations hold under $\cH_0$, with $p_d$ replaced by $p_f$.

Returning to~(\ref{eq:rs}), 
let us assume first that $\cH_1$ is in force and consider the following approximation. For any positive integer~$\omega_k$:
\begin{align}
\bz_{k\ell} &= (1-\eta_k) \sum_{i=1}^{\omega_k} \eta_k^{i-1} \bb_\ell(i) + (1-\eta_k)\sum_{i=\omega_k+1}^{\infty} \eta_k^{i-1} \bb_\ell(i) \nonumber \\
&\approx (1-\eta_k) \sum_{i=1}^{\omega_k} \eta_k^{i-1} \bb_\ell(i) + \eta_k^{\omega_k} \; \; \dfz \; \; \widehat \bz_{k\ell},
\label{eq:basic}
\end{align}
where the approximation consists of assuming that all the binary digits ${\bb_\ell(\omega_k+1)}, {\bb_\ell(\omega_k+2)}, \dots$ are equal to the most likely value $+1$, yielding
$(1-\eta_k)\sum_{i=\omega_k+1}^{\infty} \eta_k^{i-1} \bb_\ell(i)=\eta_k^{\omega_k}$. To the other extreme, when they are all equal to the most unlikely value $-1$, we have 
$(1-\eta_k)\sum_{i=\omega_k+1}^{\infty} \eta_k^{i-1} \bb_\ell(i)=-\eta_k^{\omega_k}$, which shows that the error involved in the approximation~(\ref{eq:basic}) is bounded (with probability one) by
\beq
0\le \widehat \bz_{k\ell}-\bz_{k\ell} \le 2 \, \eta_k^{\omega_k},
\label{eq:basic2}
\eeq
with the upper bound achieved when ${\bb_\ell({\omega_k}+1)}={\bb_\ell({\omega_k}+2)}= \cdots =-1$.
Note, from~(\ref{eq:basic}), that we have approximated the random variable $\bz_{k \ell}$ by a discrete random variable $\widehat \bz_{k \ell}$ taking on $2^{\omega_k}$ values. 

To choose the value of ${\omega_k}$, we need to see the effect of the approximation on the variable $\bz_k^\star$. 
Let $\widehat \bz_k^\star$ be the approximate version of $\bz_k^\star$, obtained when 
$\bz_{k\ell}$ is replaced by $\widehat \bz_{k\ell}$ in~(\ref{eq:begin2}).
Using~(\ref{eq:basic2}) in~(\ref{eq:begin2}), we have 
\begin{align}
0 \le \widehat \bz_k^\star - \bz_k^\star &\le (\E_1 \bx-\E_0 \bx) (1-a_k) \frac{\eta_k^{\omega_k}}{1-\eta_k} \nonumber \\
& \le (\E_1 \bx-\E_0 \bx) \frac{\eta_k^{\omega_k}}{1-\eta_k}.
\label{eq:err}
\end{align}
To control the error in the approximation, we enforce the condition 
\beq
\widehat \bz_k^\star - \bz_k^\star \le (\E_1 \bx-\E_0 \bx) \frac{\eta_k^{\omega_k}}{1-\eta_k} \le \epsilon_{k,h}
\label{eq:epk}
\eeq
for some ``small'' $\epsilon_{k,h}>0$, whose choice will be discussed later. 
Therefore, the index ${\omega_k}$ introduced in~(\ref{eq:basic}) is selected to comply with~(\ref{eq:epk}) as follows:
\beq
{\omega_k} =\left \lceil \frac{\log \frac{\E_1 \bx-\E_0 \bx}{\epsilon_{k,h}(1-\eta_k)}}{\log \frac{1}{\eta_k}} \right \rceil ,
\label{eq:r}
\eeq
where $\lceil z \rceil$ is the smallest integer $ \ge z$.  
Note that the right-hand side of~(\ref{eq:r}) is a decreasing function of $\epsilon_{k,h}$ and an increasing function of $\eta_k$, when $\epsilon_{k,h} < (\E_1 \bx-\E_0 \bx)/(1-\eta_k)$. Note also, from~(\ref{eq:r}), that $\omega_k$ depends also on the hypothesis $\cH_h$, $h=0,1$. This dependence is not made explicit for simplicity of notation.

\subsubsection{First-Order Approximation of $\widehat \bz_{k\ell}$}
\label{sec:foap}
We now exploit the assumption $p_d \approx 1$ to get an approximation for $\widehat \bz_{k\ell}$. 
The value taken by the random variable $\widehat \bz_{k\ell}$ in~(\ref{eq:basic}) depends on the value of the binary variables $\bb_\ell(1),\dots,\bb_\ell({\omega_k})$. For large values of $p_d$, such string of ${\omega_k}$  binary variables typically consists of many ``$+1$'' and a few ``$-1$''. This suggests to quantize  
the random variable $\widehat \bz_{k\ell}$ taking on $2^{\omega_k}$ values, into a random variable that takes on only the $({\omega_k}+1)$ values corresponding to strings with at most one ``$-1$''.
In Table~\ref{tab:firstoa} we report, arranged in ascending order, the $({\omega_k}+1)$ values taken by the quantized $\widehat \bz_{k\ell}$, followed by the string of the ${\omega_k}$ binary digits that generates that value, which is referred to as the \emph{pattern} (``$+$'' means ``$+1$'', and ``$-$'' means ``$-1$''). In Table~\ref{tab:firstoa} the subindex $k$ to $\eta_k$ and $\omega_k$ is omitted.
\begin{table}[h]
\caption{Random variable $\widehat \bz_{k\ell}$ for the first-order approximation.}
\centering 
\begin{tabular}{c|c|c}
\hline \hline 
value & pattern & probability  \\
\hline
$1-2(1-\eta)$  & $- + +  + \dots  +  +$  & $(1-p_d)$  \\
$1-2\eta(1-\eta)$  & $+ - +  + \dots  +  +$ & $(1-p_d) p_d$  \\
$1-2\eta^2(1-\eta)$  & $+ + -  + \dots +  +$ & $(1-p_d) p_d^2$  \\
$\vdots$ & $\vdots$   & $\vdots$ \\ 
$1-2\eta^{{\omega}-2}(1-\eta)$ & $+ + +  + \dots  -  +$ & $ (1-p_d) p_d^{{\omega}-2} $ \\
$1-2\eta^{{\omega}-1}(1-\eta)$ & $+ + +  + \dots  +  -$ & $(1-p_d) p_d^{{\omega}-1} $\\
$1$ & $+ + +  +  \dots  +  + $ & $p_d^{\omega}$ \\
\hline \hline
\end{tabular}
\label{tab:firstoa}
\end{table}

We do not implement a standard quantizer. Instead, the probabilities assigned to the $({\omega_k}+1)$ values are shown in the last column of Table~\ref{tab:firstoa}, and are computed as follows. 
The probability assigned to the first pattern ``$- + + + +\dots+$'' is the sum of the probabilities of all patterns in the form 
``$-  \star \star \dots\star$'', where ``$\star$'' can be either ``$+$'' or ``$-$''. The probability of the second pattern  
``$+ - + + + \dots+$'' is the sum of the probabilities of all patterns in the form 
``$+ -  \star \dots\star$'', and so forth. The general rule is to replace by stars all the symbols in the pattern following the symbol ``$-$'', 
if any.

Other approximations can be conceived. First, to the values shown in the first column of Table~\ref{tab:firstoa}, one could assign a probability proportional to that of the corresponding pattern. But this approach amounts to neglect the values of $\widehat \bz_{k\ell}$ having small probability, which leads to a poor approximation. More appealing would be to implement a regular quantization of $\widehat \bz_{k\ell}$, with quantization regions that are intervals. However, for $\eta_k> 1/2$, it may not be easy to identify the sequences $\bb_\ell(1),\dots,\bb_\ell({\omega_k})$ that, inserted in~(\ref{eq:basic}), yield a value belonging to a prescribed interval. 
The approach followed in Table~\ref{tab:firstoa}, instead, is analytically straightforward for any value of $\eta_k$ and, for $\eta_k\le 1/2$, just amounts to a regular quantization of the random variable $\widehat \bz_{k\ell}$, as it can be shown by simple algebra.

\subsubsection{Second-Order Approximation of $\widehat \bz_{k\ell}$}
\label{sec:soap}
Along the same lines of the approximation just developed, one can assume that the sequence $\bb_\ell(1),\dots,\bb_\ell({\omega_k})$ in~(\ref{eq:basic}) contains at most \emph{two} occurrences of the unlikely digit ``$-1$'', 
in which case the random variable $\widehat \bz_{k\ell}$ is approximated by a random variable that takes on the $1+{\omega_k}+{\omega_k}({\omega_k}-1)/2$ values shown in Table~\ref{tab:secondoa}, with associated patterns and probabilities. In Table~\ref{tab:secondoa} the subindex $k$ to $\eta_k$ and $\omega_k$ is omitted.

\begin{table}[h]
\caption{Random variable $\widehat \bz_{k\ell}$ for the second-order approximation.}
\centering 
\scriptsize
\begin{tabular}{c|c|c}
\hline \hline 
value & pattern & probability  \\
\hline
$1-2(1-\eta)-2\eta(1-\eta)$ & $- - +  + \dots +  +  +$ & $(1-p_d)^2 $  \\ 
$1-2(1-\eta)-2\eta^2(1-\eta)$ & $- + -  + \dots +  +  +$ &$(1-p_d)^2 p_d $ \\
\vdots & \vdots &  \vdots \\ 
$1-2(1-\eta)-2\eta^{{\omega}-1}(1-\eta)$ & $- + +  + \dots +  +  - $ &$(1-p_d)^2 p_d^{{\omega}-2}$  \\
$1-2(1-\eta)$   & $ -  +  +   +  \dots  +   +   + $ &   $(1-p_d) p_d^{{\omega}-1}$  \\
$1-2\eta(1-\eta)-2\eta^2(1-\eta)$   &$+  -  -   +  \dots  +   +   + $&   $(1-p_d)^2 p_d$  \\
$1-2\eta(1-\eta)-2\eta^3(1-\eta)$   &$+  -  +   -  \dots  +   +   + $&   $(1-p_d)^2 p_d^2$  \\
\vdots & \vdots &  \vdots \\ 
$1-2\eta(1-\eta)-2\eta^{{\omega}-1}(1-\eta)$   &$+  -  +   +  \dots  +   +   - $&   $(1-p_d)^2 p_d^{{\omega}-2}$  \\
$1-2\eta(1-\eta)$   &$+  -  +   +  \dots  +   +   + $&   $(1-p_d) p_d^{{\omega}-1}$  \\
$1-2\eta^2(1-\eta)-2\eta^3(1-\eta)$   &$+  +  -   -  \dots  +    +   + $&   $(1-p_d)^2 p_d^2$  \\
\vdots & \vdots &  \vdots \\ 
$1-2\eta^2(1-\eta)-2\eta^{{\omega}-1}(1-\eta)$   &$+  +  -   +  \dots  +   +   - $&   $(1-p_d)^2 p_d^{{\omega}-2}$  \\
$1-2\eta^2(1-\eta)$   &$+  +  -   +  \dots  +    +   + $&   $(1-p_d) p_d^{{\omega}-1}$  \\
\vdots & \vdots &  \vdots \\ 
\vdots & \vdots &  \vdots \\ 
\hspace*{-5pt}$1-2\eta^{{\omega}-3}(1-\eta)-2\eta^{{\omega}-2}(1-\eta)$ \hspace*{-7pt}  &$+  +  +   +  \dots  -   -   + $&   $(1-p_d)^2 p_d^{{\omega}-1}$  \\
\hspace*{-5pt}$1-2\eta^{{\omega}-3}(1-\eta)-2\eta^{{\omega}-1}(1-\eta)$ \hspace*{-7pt}   &$+  +  +   +  \dots  -   +   - $&   $(1-p_d)^2 p_d^{{\omega}-2}$  \\
\hspace*{-5pt}$1-2\eta^{{\omega}-3}(1-\eta)$  \hspace*{-7pt} &$+  +  +   +  \dots  -   +   + $&   $(1-p_d) p_d^{{\omega}-1}$  \\
\hspace*{-5pt}$1-2\eta^{{\omega}-2}(1-\eta)-2\eta^{{\omega}-1}(1-\eta)$ \hspace*{-7pt}  &$+  +  +   +  \dots  +   -   - $&   $(1-p_d)^2 p_d^{{\omega}-2}$  \\
\hspace*{-5pt}$1-2\eta^{{\omega}-2}(1-\eta) $ \hspace*{-7pt} &$+  +  +   +  \dots  +   -   + $&   $(1-p_d) p_d^{{\omega}-1}$  \\
\hspace*{-5pt}$1-2\eta^{{\omega}-1}(1-\eta)$  \hspace*{-7pt} &$+  +  +   +  \dots  +   +   - $&   $(1-p_d) p_d^{{\omega}-1}$  \\
$1$   &$+  +  +   +  \dots  +   +   + $&   $p_d^{\omega}$  \\
 \hline \hline 
\end{tabular}
\label{tab:secondoa}
\end{table}

The probabilities in the last column of Table~\ref{tab:secondoa} are computed as follows.
Consider a generic pattern with two occurrences of ``$-$'', and let us replace with the symbol  ``$\star$'' all the ``$+$'' appearing to the right of the rightmost ``$-$''. Then, the probability assigned to that pattern is the sum of the probabilities corresponding to all the distinct patterns obtained by assigning to the stars either the symbol ``$+$'' or the symbol ``$-$''. 
For the~${\omega_k}$ patterns with only one ``$-$'', the probabilities are exactly those of the pattern, without modification.
To understand why we use this convention, consider for instance the pattern ``$++-+ \dots +++$'', corresponding to the value $1-2\eta_k^2(1-\eta_k)$. All probabilities of patterns of the form ``$++-\star \dots \star  \star \star$'' are included in the probability of some pattern with two ``$-$'', except
the single pattern ``$++-+ \dots +++$'', whose probability is just $(1-p_d)p_d^{{\omega_k}-1}$.

Straightforward algebra shows that the values in the first column of Table~\ref{tab:secondoa} are arranged in ascending order only if\footnote{Note that $\frac{\sqrt{5}-1}{2}$ is the \emph{golden ratio} $\frac{\sqrt{5}+1}{2}$ minus one. The golden ratio is the ratio of successive terms in the Fibonacci series~\cite{schroeder}.}  $\eta_k < \frac{\sqrt{5}-1}{2} \approx 0.618$, but the order of the elements does not matter for the final approximation. Furthermore, depending on the values of $\eta_k$, it may happen that several entries in the first column of Table~\ref{tab:secondoa} are closer than $2 \eta_k^{\omega_k}$, see~(\ref{eq:basic2}). In this case, it may be convenient 
to reduce the cardinality of the random variable $\widehat \bz_{k\ell}$ by aggregating these values in a single value and associating with it the sum of the probabilities of the merged entries.

So far, we have developed the approximation under the assumption that hypothesis $\cH_1$ is in force. Exploiting the problem symmetry, it can be seen that the approximation under $\cH_0$ can be obtained by applying the procedure described for the hypothesis $\cH_1$ to the random variable $-\bz_{k\ell}$, and by replacing $p_d$ with $(1-p_f)$.

\subsubsection{Approximate Distribution for $\bz_k(\infty)$}
\label{sec:adz}

As shown in~(\ref{eq:basic}), we replace $\bz_{k\ell}$ by the discrete random variable $\widehat \bz_{k\ell}$ with $2^{\omega_k}$ values, where ${\omega_k}$ is given in~(\ref{eq:r}). In turn, $\widehat \bz_{k\ell}$ is approximated by the lower-cardinality discrete random variable shown in Table~\ref{tab:firstoa}, if the first-order approximation is used, and shown in Table~\ref{tab:secondoa} for the more accurate second-order approximation. In the latter case, the alphabet of the random variable is further reduced by grouping together values that are closer than $2\eta_k^{\omega_k}$. This way, an approximate  
 probability mass function (PMF) of  $\bz_{k\ell}$ is obtained.

As indicated in~(\ref{eq:begin2}), the random variable $\bz_{k\ell}$ must be multiplied by $(\E_1 \bx-\E_0 \bx)/2$, then added to $(\E_1 \bx+\E_0 \bx)/2$, and finally multiplied by $c_{k\ell}/(1-\eta_k)$. 
The resulting random variables represent the individual summands of the sum over $\ell=1,\dots, S$, appearing in~(\ref{eq:begin}).
Note that these random variables are independent.
Summing over $\ell=1,\dots, S$, corresponds to convolving the PMFs of the summands, which finally provides the desired PMF [equivalently, the CDF $F_{\bz,h}(u)$] of the steady-state discrete contribution $\bz_k(\infty)$. 
The number of convolutions is $|{\cal N}_k|-2$, and these convolutions can be easily implemented numerically, but the computation might become cumbersome when the alphabet of the individual PMFs becomes excessively large. 
Thus, after computing each convolution, similarly to what we have done before, the values of the resulting variable that are close to each other more than $(\E_1 \bx- \E_0 \bx) \eta_k^{\omega_k}/(1-\eta_k) $ [see~(\ref{eq:epk})] are merged  into a single value, and the corresponding probabilities are summed up.

Summarizing, we have derived simple expressions for the distribution of the random variables~$c_{k\ell}\sum_{i=1}^{\infty}  \eta_k^{i-1} \, \widetilde \bx_{\ell}(i)$, $\ell=1,\dots,S$, in~(\ref{eq:begin}). Each random variable has been approximated by a discrete random variable whose PMF has been characterized analytically. The desired PMF for $\bz_k(\infty)$ is 
the PMF of the finite sum $\sum_{\ell=1}^S c_{k\ell}\sum_{i=1}^{\infty}  \eta_k^{i-1} \, \widetilde \bx_{\ell}(i)$, and is obtained by computing numerically the convolutions that correspond to the sum over~$\ell$.
For later use, let us denote this PMF as follows:
\beq
\P_h \left [\bz_k(\infty)= z_{i,h} \right]=\nu_{i,h} \, , \quad i=1,\dots,L,
\label{eq:PMF}
\eeq
where $L \ll 2^{{\omega_k}}$. In other words, the CDF $F_{\bz,h}(z)$, $z \in \Re$, of $\bz_k(\infty)$ is approximated by a 
staircase function with steps of size $\{\nu_{i,h}\}$ at the $L$ points $\{z_{i,h}\}$.

An informal discussion is useful to better understand the role of the system parameters in determining the distribution of the discrete component $\bz_k(\infty)$. Consider first hypothesis~$\cH_1$. Since the typical values of the step-size $\mu$ are much less than unity, let us assume $\mu=0.1$. For $a_k$ very small so that $\eta_k \ll 1$, we see from Tables~\ref{tab:firstoa} and~\ref{tab:secondoa} that the random variable $\widehat \bz_{k\ell}$ takes value $1$ with probability close to one, and value $\approx -1$ with probability close to zero. 
If node~$k$ has a single neighbor, i.e., $|\N_k|=2$, the sum in~(\ref{eq:begin2}) disappears and the discrete component $\bz_k(\infty)$ takes only two values with non-negligible probability: $\approx \frac{1-a_k}{1-\eta_k} \E_1 \bx \approx \E_1 \bx$ with large probability, and $\approx \frac{1-a_k}{1-\eta_k} \E_0 \bx \approx  \E_0 \bx$. 
In this situation, moderate variations of $\mu$ have little effect on $\eta_k$, and hence a little effect at all. When~$a_k$ grows, so does~$\eta_k$, and 
from Tables~\ref{tab:firstoa} and~\ref{tab:secondoa} we see that the PMF of $\bz_k(\infty)$ is enriched by additional contributions 
at points between the extremes, and these extremes become closer to each other. In this case, diminishing the step-size~$\mu$ makes~$\eta_k$ 
larger, which tends to further increasing the number of points of the PMF (the effect on the position of the extreme values is modest, unless $a_k$ is very large). 
For nodes with larger connectivity $|\N_k|$, because of the sum in~(\ref{eq:begin2}), in each of the above situations,
the PMF of the discrete component is enriched by the contributions due to the convolutions.
Similar arguments apply under~$\cH_0$.

\subsection{Distribution of the Steady-State Statistic}
\label{sec:DSSS}

The CDF of $\bu_k(\infty)$ defined in~(\ref{eq:uinf}) is $F_{\bu,h}(u)$, and the CDF of $\bz_k(\infty)$ defined in~(\ref{eq:zinf}) is $F_{\bz,h}(z)$. We have seen in the previous sections how to compute approximately these two distributions, and we now discuss how to obtain the CDF $F_{\by,h}(y)$, $y \in \Re$, of the variable 
\beq
\by_k(\infty) \dfz \lim_{n\rightarrow \infty} \by_k(n)= \bu_k(\infty) + \bz_k(\infty)
\eeq 
which characterizes the state of node $k$ in the steady-state regime of $n\rightarrow \infty$.
We assume, as already done in Theorem~1, that $F_{\bu,h}(u)$ admits a density, denoted by $f_{\bu,h}(u)$. 
This implies that $F_{\by,h}(y)$ also admits a density, and
the CDF $F_{\by,h}(y)$ is given by the convolution~\cite[pp.\ 144--146]{FellerBookV2} 
\beq
F_{\by,h}(y)= \int_{-\infty}^\infty  F_{\bz,h}(\xi)  \, f_{\bu,h}(y-\xi) \, d\xi.
\label{eq:conv}
\eeq
Using approximation~(\ref{eq:basic}) in Sec.~\ref{sec:z}, the random variable $\bz_k^\star$ is replaced by 
$\widehat \bz_k^\star$ [see~(\ref{eq:err})], whose CDF will be denoted by $\widehat F_{\bz,h}(z)$, $z \in \Re$. From condition~(\ref{eq:epk}) the following relationship between events follows
\begin{align}
&\{ \bz_k^\star \le z-\epsilon_{k,h} \} \subseteq \{\, \widehat \bz_k^\star \le z \}  \subseteq \{\bz_k^\star \le z\},
\end{align}
implying that
\beq
F_{\bz,h}(z-\epsilon_{k,h}) \le \widehat F_{\bz,h}(z) \le  F_{\bz,h}(z).
\label{eq:CDFs}
\eeq

Before applying the approximations described in Secs.~\ref{sec:foap} and~\ref{sec:soap},  the random variable $\widehat \bz_k^\star$ is discrete and takes on (at most) $2^{\omega_k (|\N_k|-1)}$ values. For $i=1,\dots,2^{\omega_k (|\N_k|-1)}$, let $\{\widehat z_{i,h}\}$ denote these values and $\{\widehat \nu_{i,h}\}$ be the corresponding probabilities, so that
$\widehat F_{\bz,h}(z)$ is a staircase function with steps of size $\{\widehat \nu_{i,h}\}$ at the $2^{\omega_k (|\N_k|-1)}$ points $\{\widehat z_{i,h}\}$. From~(\ref{eq:conv}) and ~(\ref{eq:CDFs}):
\begin{align}
&F_{\by,h}(y-\epsilon_{k,h})=\int_{-\infty}^\infty  F_{\bz,h}(\xi)  \, f_{\bu,h}(y-\epsilon_{k,h}-\xi) \, d\xi \nonumber \\
&\quad = \int_{-\infty}^\infty  F_{\bz,h}(\xi-\epsilon_{k,h})  \, f_{\bu,h}(y-\xi) \, d\xi  \nonumber \\
&\quad \le \int_{-\infty}^\infty  \widehat F_{\bz,h}(\xi)  \,  f_{\bu,h}(y-\xi) \, d\xi  = \sum_i \widehat \nu_{i,h} F_{\bu,h}(y-\widehat z_{i,h})  \nonumber \\
&\quad \le \int_{-\infty}^\infty  F_{\bz,h}(\xi)  \,  f_{\bu,h}(y-\xi) \, d\xi = F_{\by,h}(y).
\end{align}
This proves that 
\beq
F_{\by,h}(y) \approx \sum_{i=1}^{2^{\omega_k (|\N_k|-1)}} \widehat \nu_{i,h} F_{\bu,h}(y-\widehat z_{i,h})
\label{eq:finapp2}
\eeq
provided that
\beq
\int_{-\infty}^\infty  \hspace*{-5pt} F_{\bz,h}(\xi)  \, f_{\bu,h}(y-\epsilon_{k,h}-\xi) \, d\xi
\approx
\int_{-\infty}^\infty  \hspace*{-5pt}  F_{\bz,h}(\xi)  \,  f_{\bu,h}(y-\xi) \, d\xi,
\label{eq:smoothness}
\eeq
which is true when $f_{\bu,h}(u)$ is sufficiently smooth over intervals of length $\epsilon_{k,h}$.
The value of $\epsilon_{k,h}$ will be chosen just to ensure this condition, see Sec.~\ref{sec:compe}.

The final approximation of the steady-state variable $\by_k(\infty)$ is obtained from~(\ref{eq:finapp2}) after
manipulating the random variable~$\widehat \bz_k^\star$ as detailed in Secs.~\ref{sec:foap}, \ref{sec:soap}, and \ref{sec:adz},
which reduces its cardinality from $2^{\omega_k (|\N_k|-1)}$ to the much smaller value $L$. This yields
\beq
F_{\by,h}(y) \approx \sum_{i=1}^{L}  \nu_{i,h} \, F_{\bu,h}(y- z_{i,h}), \qquad y \in \Re,
\label{eq:finapp}
\vspace*{-3pt}
\eeq
where $\{\nu_{i,h}\}$ and $\{z_{i,h}\}$ are defined in~(\ref{eq:PMF}) and the sum involves $L \ll 2^{\omega_k  (|\N_k|-1)}$ terms.
Our final approximation~(\ref{eq:finapp}) is only semi-analytical, in the sense that we have an analytical expression for $F_{\bu,h}(u)$ but this expression involves the truncation of series to finite sums, see~(\ref{eq:appnew2}). Similarly, after analytical derivations, computing the sequences $\{\nu_{i,h}\}$ and $\{z_{i,h}\}$ requires numerical convolutions (over discrete, low-cardinality, sets). In both cases, the numerical procedures are very simple.

The role of the system parameters on the steady-state distribution can be understood by recalling the comments made at 
the end of Sec.~\ref{sec:adz}, and exploiting expression~(\ref{eq:finapp}). From~(\ref{eq:finapp}) we see that the CDF of $\by_k(\infty)$
consists of the superposition of $L$ copies of the CDF of the continuous component $\bu_k(\infty)$, scaled by $\nu_{i,h}$ and centered at the points $z_{i,h}$, $i=1,\dots,L$.
Assuming that $f_{\bu,h}(u)$ is unimodal, the interplay between its width and the spacing between the $z_{i,h}$'s determines the smoothness of the CDF of $\by_k(\infty)$.
Quantitative predictions require a case-by-case analysis, but for many scenarios of practical interest, as those presented in Sec.~\ref{sec:compe}, 
the general trend is as follows. Suppose $\mu=0.1$, $|\N_k|=2$, and $a_k \ll 1$. 
As $|\N_k|$ and/or $a_k$ grow, the CDF $F_{\by,h}(y)$ under $\cH_0$ (resp.\ $\cH_1$) moves progressively from a jump-wise shape with larger jumps for small (resp.\ large) values of $y$, to a jumpless smoother shape. This behavior will be confirmed in Sec.~\ref{sec:compe}.

The theory developed so far allows us to approximate the distribution $F_{\by,h}(y)$ of the steady-state statistic $\by_k(\infty)$ for most values of $\mu$ and $a_k$ of interest.  However, for $\eta_k \to 1$ the approach may become problematic, for several reasons. From a numerical viewpoint,
when $\eta_k$ is too close to unity the index~${\omega_k}$ appearing in~(\ref{eq:r}) may be excessively large. More important, for $\eta_k \to 1$ we see from~(\ref{eq:di}) that
the dispersion index of the continuous component $\bu_k(\infty)$ converges to zero. In addition, $\eta_k \to 1$ is equivalent to $\mu \to 0$ and $a_k\to 1$,
and the expected value $\E_h[\bu_k(\infty)]$ of the continuous component takes different values depending on the order of the two limits, see the first equation in~(\ref{eq:mev}). The same indeterminateness may affect the discrete component $\bz_k(\infty)$, because of the factor $\frac{1-a_k}{1-\eta_k}$ appearing in~(\ref{eq:begin2}) when $|\N_k|=2$.
Thus, $\eta_k \to 1$ is a challenging scenario in which a separate analysis of the discrete and continuous components may not be sufficient. A different approach is required, 
which is presented in the next section.

\subsection{Asymptotic Normality}

Ignoring the transient term, the random process $\by_k(n)$ in~(\ref{eq:rec2}) can be expressed as the sequence of partial sums 
\beq
\by_k(n)= \sum_{i=1}^n \bw_k(n,i)
\eeq
of a \emph{triangular array} $\{\bw_k(n,i)\}$, with $i \le n$, and $n=1,2,\dots$, where\footnote{The quantity $\bw_k(n,i)$ is not to be confused with $\bw_k^\star$ 
defined in~(\ref{eq:winf}).} 
\begin{align}
\bw_k(n,i) \dfz \eta_k^{i-1} \Big [ \mu a_k \bx_k(n-i+1) + \sum_{\ell=1}^S c_{k\ell} \widetilde \bx_{\ell} (n-i+1) \Big ].
\label{eq:array}
\end{align}
Clearly, it holds that:
\begin{align}
\E_h[\bw_k(n,i)]&=  \eta_k^{i-1} \left [ \mu a_k \E_h \bx + (1-a_k) \, \E_h \widetilde \bx  \right ] ,
\label{eq:Ehfirst} \\
\V_h[\bw_k(n,i)]& = \eta_k^{2(i-1)} \Bigg (\mu^2 a_k^2 \,  \V_h \bx + \V_h \widetilde \bx  \, \sum_{\ell \neq k} a_{k\ell}^2\Bigg ),
\label{eq:Vhfirst}
\end{align}
where~(\ref{eq:Vhfirst}) follows by using $c_{kk}=0$ in~(\ref{eq:array}) --- see~(\ref{eq:c}).
It is worth noting that the two sequences $\E_h[\bw_k(n,i)]$ and $\V_h[\bw_k(n,i)]$ do not depend on the index $n$: they are standard sequences, not arrays [the same is true for any moment of $\bw_k(n,i)$, because the $\bx$'s  are i.i.d.]. 
From~(\ref{eq:Ehfirst}) and~(\ref{eq:Vhfirst}):
\begin{align}
m_{k,h}(n) &\dfz \E_h[\by_k(n)] = \sum_{i=1}^n \E_h[\bw_k(n,i)] \nonumber \\
& = \frac{1-\eta_k^n}{1-\eta_k}  \left [\mu a_k \E_h \bx + (1-a_k) \,\E_h \widetilde \bx \right ] ,\label{eq:dfzm} \\
s_{k,h}^2(n) & \dfz \V_h[\by_k(n)] =\sum_{i=1}^n \V_h[\bw_k(n,i)] \nonumber \\
&=\Bigg (\mu^2 a_k^2 \V_h \bx +  \V_h \widetilde \bx \, \sum_{\ell \neq k} a_{k\ell}^2\Bigg ) \frac{1-\eta_k^{2n}}{1-\eta_k^{2}}.
\label{eq:dfzs}
\end{align}
The network topology plays its role in $m_{k,h}(n)$ and $s_{k,h}^2(n)$
through the quantities $S$, $\{a_k\}_{k=1}^S$, and $ \sum_{\ell \neq k} a_{k\ell}^2$. As to the latter, we have 
\beq
\frac{(1-a_k)^2}{S-1} \le \sum_{\ell \neq k} a_{k\ell}^2 \le (1-a_k).
\label{eq:bounds}
\eeq
To prove~(\ref{eq:bounds}), note that $a_k \le 1$ implies $\sum_{\ell \neq k} a_{k\ell}^2$ $\le \sum_{\ell \neq k} a_{k\ell}$ $= (1-a_k)$, and the upper bound follows.
The lower bound can be derived by applying the Cauchy-Schwarz inequality to the sequence $\{{a_{k\ell}}\}$ and to the constant sequence of all ones, which yields $(\sum_{\ell \neq k} a_{k\ell})^2\le$ $(S-1)$ $\sum_{\ell \neq k} a_{k\ell}^2$.
Of special interest is the situation in which all the neighbors of a given node are equally \emph{reliable}, and there is no reason to assign different weights to different neighbors. In this case it makes sense to impose that the off-diagonal entries on each row of matrix~$A$ are equal. Then, the lower bound $ \sum_{\ell \neq k} a_{k\ell}^2=\frac{(1-a_k)^2}{S-1}$ in~(\ref{eq:bounds}) is achieved, and the only topological parameters are the number of nodes $S$ and the self-combination coefficients $a_k$.

The limiting quantities $m_{k,h}(\infty) \dfz \lim_{n \rightarrow \infty} m_{k,h}(n)$ and $s_{k,h}^2(\infty) \dfz \lim_{n \rightarrow \infty} s_{k,h}^2(n)$ are immediately obtained from~(\ref{eq:dfzm}) and~(\ref{eq:dfzs}) by noting that $\lim_{n \rightarrow \infty}\eta_k^n=0$, and we have the following result.

\vspace*{5pt}
\noindent \textbf{Theorem 2 (Asymptotic normality):} 
\emph{Suppose that there exists $\beta>0$ such that $\E |\bx|^{2+\beta}  < \infty$. Then, under hypothesis $\cH_h$, $h=0,1$, we have:
\begin{align} 
\lim_{\eta_k\to1} \frac{\by_k(\infty)-m_{k,h}(\infty)}{s_{k,h}(\infty)} = \bg,
\label{eq:CLT}
\end{align}
where $\bg$ is a standard Gaussian random variable, and the convergence is in distribution.}
\vspace*{3pt}

\noindent \emph{\bf Proof.} The result essentially follows from the CLT for triangular arrays under the Lyapunov condition, see e.g.,~\cite[Th. 27.3]{billingsley-book2}. One key modification in the argument  is that the limit involves also the continuous parameter~$\eta_k$. 
The detailed proof is given in Appendices~\ref{app:theo2} and~\ref{app:lyapo}.\hfill$\square$
\vspace*{3pt}

Note that, for the theorem to hold, it suffices that $\E |\bx|^{3}$ is finite, a condition that is usually easily checked.
It is also worth noting that $\mu\rightarrow 0$ is not sufficient to ensure the asymptotic normality of (a normalized version of) $\by_k(\infty)$.
Technically, the reason is that the Lyapunov condition~(\ref{eq:lyapo2}) in Appendix~\ref{app:lyapo} does not hold if one replaces the limit $\eta_k \rightarrow 1$ with the limit $\mu\rightarrow 0$ --- see the arguments at the end of Appendix~\ref{app:lyapo}.
The discussion below elaborates on this issue from a more intuitive perspective.

There are several substantial differences between the classical diffusion scheme~(\ref{eq:old1})-(\ref{eq:old2}) studied in~\cite{BracaetalIT,MattaSIPN16}, and the version~(\ref{eq:new1})-(\ref{eq:new2}) proposed here for one-bit messaging.
One critical difference is just the regime in which the steady-state statistic is normally distributed. For~(\ref{eq:old1})-(\ref{eq:old2}), the steady-state statistic is the rightmost sum in~(\ref{eq:recold}). 
From that expression we see that the step size $\mu$ appears as a common factor for the $n$ elements of the sum. For $n\rightarrow \infty$, this implies that  the steady-state statistic is the sum of infinitely many infinitesimal contributions, provided that $\mu \rightarrow 0$. The CLT applies just to that kind of situations where a large number of independent contributions are added together and none of them dominates the other terms, namely the individual variances of the components are small in comparison to their  sum.   

Contrast this situation with the case of one-bit messaging in~(\ref{eq:rec}). When $n\rightarrow \infty$, the infinite sum appearing in the second line of equation~(\ref{eq:rec}) consists of infinitesimal contributions when $\mu\rightarrow 0$. However, this is not the case for the infinite sum appearing in the third line of~(\ref{eq:rec}), which takes into account the decisions of neighboring nodes. In order to invoke the CLT, we must impose that the contributions of this latter infinite sum become vanishing, and this requires $c_{k\ell} \rightarrow 0$, for all $k$ and $\ell$, which can be obtained by $a_k \rightarrow 1$. As a matter of fact, in~(\ref{eq:old1})-(\ref{eq:old2}) making the step size vanishing ensures the Gaussianity of the statistic, while in~(\ref{eq:new1})-(\ref{eq:new2}) we need $\eta_k=(1-\mu)a_k \rightarrow 1$, i.e., we need a vanishing step size and a combination matrix~$A$ with highly dominant diagonal entries.

So far, we have derived the approximate CDF $F_{\by,h}(y)$, $y \in \Re$, of the steady-state value $y_k(\infty)$ of node $k=1,\dots,S$,
under both hypotheses $\cH_0$ and $\cH_1$. This has been done for general values of the parameters $\mu$ and $a_k$ in the previous subsections, and for the case $\eta_k\rightarrow 1$ in the present subsection. In the next section we apply the developed theory to a couple of examples of practical relevance, and investigate the detection properties of the network, which is our final goal.

\begin{figure}
\centering 
\includegraphics[width =200pt]{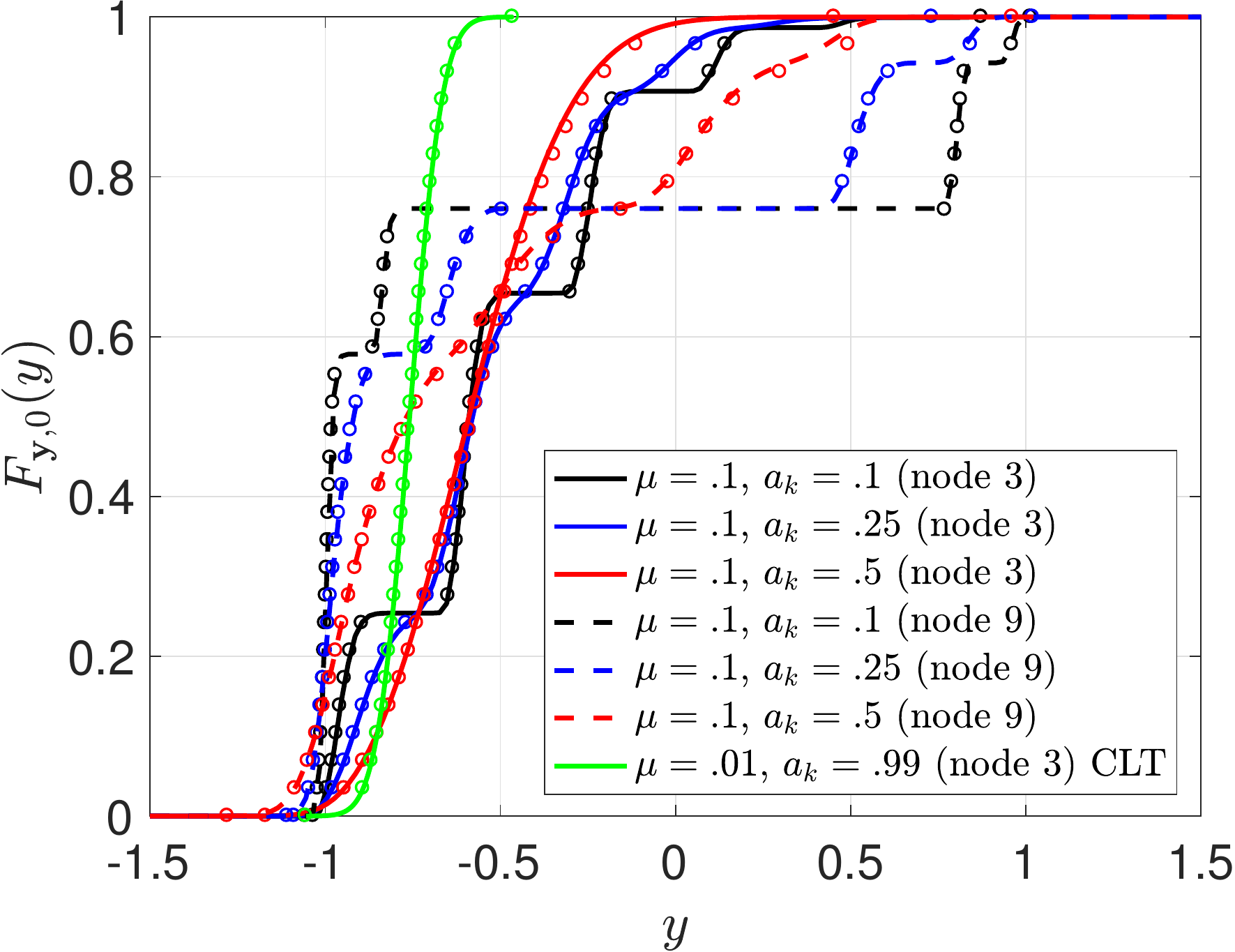}
 \caption{Gaussian example under hypothesis $\cH_0$, with $\rho=1$. It is shown the CDF $F_{\by,0}(y)$ of the steady-state variable $\by_k(\infty)$ for node $k=3$ (solid curves) and $k=9$ (dashed), obtained by the approximations developed in this paper. The symbols ``o'' show the results of computer experiments.}
      \label{fig:gausH0}
 \end{figure}

\section{Examples} 
\label{sec:compe}

We now consider two examples in which the distributions of the observations are, respectively, Gaussian and exponential. The role of the system parameters in determining the 
distributions of the steady-state variable $\by_k(\infty)$, and hence the detection capabilities of the network, is investigated. We also report the results of computer experiments to check the accuracy of the approximated formulas provided in the previous sections.

In the examples that follow, we consider the network made of $S=10$ nodes depicted in Fig.~\ref{fig:net} (this is the same network considered in~\cite{BracaetalIT}), and we refer to the right-stochastic combination matrix $A$ with entries: 
\beq
a_{k \ell}= \left \{
\begin{array}{ll}
a_k , & \ell=k, \\
\frac{1-a_k}{|{\cal N}_k|-1}, & \ell \neq k \textnormal{ are neighbors,} \\
0, &  \textnormal{otherwise,}
\end{array}
\right .
\label{eq:lap}
\eeq
where, we recall, ${\cal N}_k$ is the set of neighbors of node $k$, including $k$ itself. 
In addition, to simplify the presentation of the results, we also assume that the self-combination coefficients $a_k$ are all equal. 

As to the computer experiments designed to simulate the network evolution, the simulation procedure is as follows. For the given network topology, at each time instant $i=1,\dots n$, the observations $\bx_k(i)$ are simulated for all nodes $k=1,\dots S$, and the update rule~(\ref{eq:new1})-(\ref{eq:new2}) is implemented. 
This is made for both hypotheses $\cH_0$ and $\cH_1$, and  the resulting state $\by_k(n)$ is stored. 
The value of $n$ is large enough to make negligible the transient part of~(\ref{eq:rec2}), and to make $\by_k(n)$ a good approximation for $\by_k(\infty)$. The computation is repeated many times according to the Monte Carlo principle, thus obtaining the empirical distribution of $\by_k(\infty)$ under $\cH_0$ and under $\cH_1$. These empirical CDFs are then compared with the theoretical CDFs obtained in the previous sections. 
The number of Monte Carlo trials for obtaining each point of the empirical CDFs is $10^4$, and the number of iterations is $n=100$, if not specified otherwise.

\begin{figure}
\centering 
\includegraphics[width =200pt]{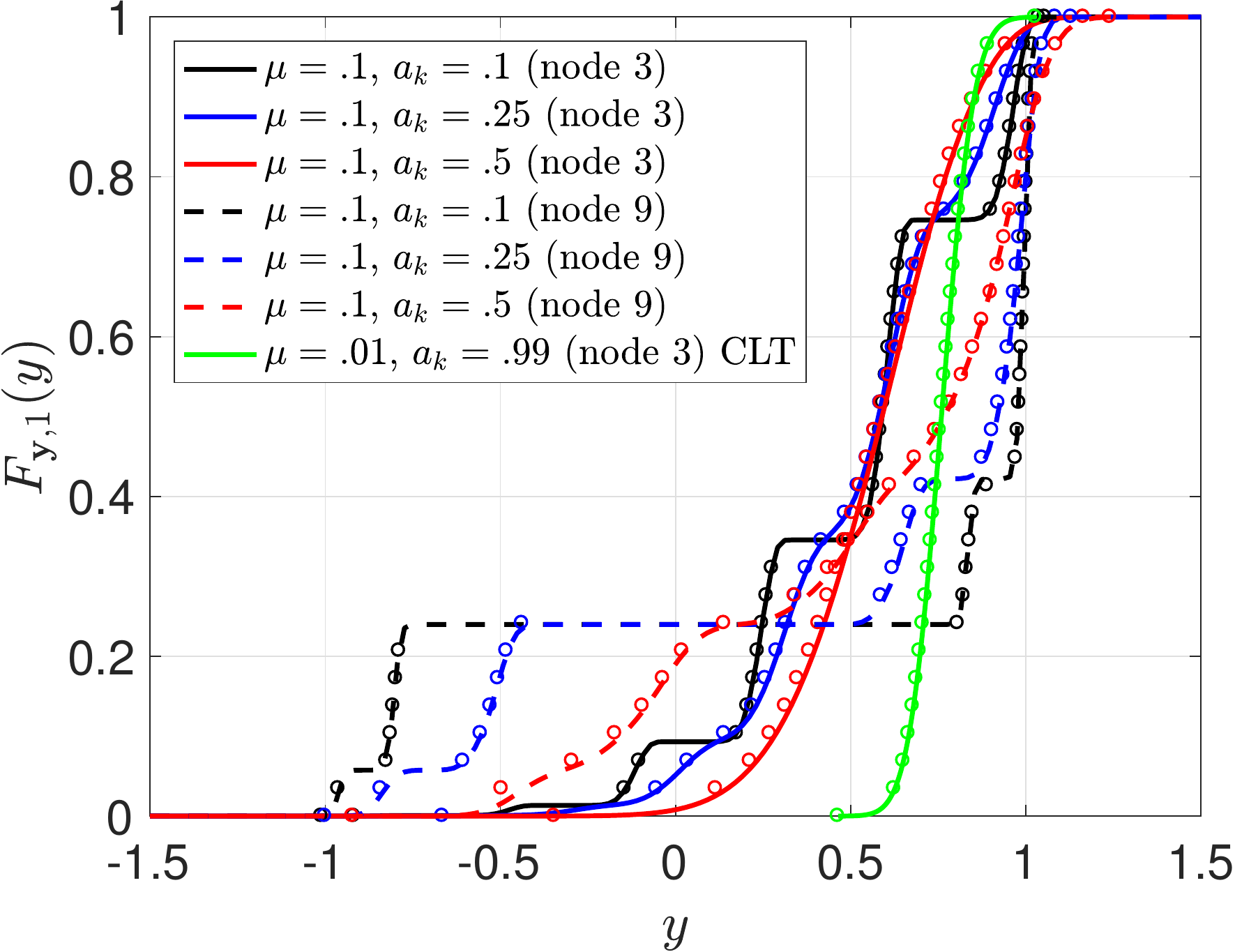}
 \caption{Gaussian example under hypothesis $\cH_1$, with $\rho=1$. It is shown the CDF $F_{\by,1}(y)$ of the steady-state variable $\by_k(\infty)$ for node $k=3$ (solid curves) and $k=9$ (dashed), obtained by the approximations developed in this paper. The symbols ``o'' show the results of computer experiments.}
      \label{fig:gausH1}
 \end{figure}

\subsection{Gaussian Observations} \label{sec:gau}
Let $\br \sim \cN(a,b)$ denote that the random variable~$\br$ is normally distributed with mean~$a$ and variance~$b$.
Suppose that the i.i.d.\ observations are so distributed: $\br \sim \cN(0,b)$ under $\cH_0$, and $\br \sim \cN(a,b)$ under $\cH_1$, with $a,b>0$. Suppose also that the nodes compute the log-likelihood ratios of these observations, according to~(\ref{eq:logli}), and the local threshold is $\gamma_{\rm loc}=0$, see~(\ref{eq:q}). Straightforward computation shows that the log-likelihood ratios are normally distributed, as follows:
\beq
\bx  \stackrel{\cH_0}{\sim} \cN(-\rho,2 \rho), \qquad \bx \stackrel{\cH_1}{\sim} \cN(\rho,2 \rho),
\eeq
where $\rho=a^2/2b$ is the Kullback-Leibler divergence from $\cH_0$ to $\cH_1$ (or vice versa: in this case the divergence is symmetric)~\cite{CT2}.

Figure~\ref{fig:gausH0} shows the CDF of $\by_k(\infty)$ under $\cH_0$, for the highly-connected node $k=3$ with $5$ neighbors, and for the weakly connected node $k=9$ with only one neighbor. In the figure we set $\rho=1$, $\mu=0.1$, and consider three different values of $a_k=0.1, 0.25, 0.5$.
The curves represent the theoretical CDFs corresponding to the second-order approximation, with solid lines for $k=3$, and dashed lines for $k=9$. 
The symbol ``o'' denotes values obtained by computer experiments. The curve in green, which refers to $\mu=0.01$ and $a_k=0.99$, shows the normal distribution predicted by Theorem~2 compared to computer simulations. In this case, $100$ iterations are not sufficient to reach the steady-state, and we use $n=1000$. 

Figure~\ref{fig:gausH1} addresses the same analysis under $\cH_1$.
It is worth mentioning that, under both hypotheses, the theoretical CDFs computed by the approximations developed in the previous section perfectly match the simulation points.

In the present example, from the remark following Theorem~1, we know that 
\beq
\bu_k(\infty) \sim \N \left(\frac{a_k \mu}{1-\eta_k} \E_h \bx ,  \frac{a_k^2 \mu^2}{1-\eta_k^2} \V_h \bx \right),
\eeq
and therefore the distribution of $\bu_k(\infty)$ is known without any approximation. 
The accuracy of the theoretical curves shown in Figs.~\ref{fig:gausH0} and~\ref{fig:gausH1} means that the (second-order) approximation developed in Sec.~\ref{sec:z} for the discrete component $\bz_k(\infty)$ is accurate. 
Note that with $\rho=1$ we have $p_d=(1-p_f) \approx 0.76$, implying that the approximation for the discrete component works quite well already for values of $p_d$ and $(1-p_f)$ only moderately close to unity.

As stated below Eq.~(\ref{eq:smoothness}), to compute the theoretical approximation of $F_{\by,h}(y)$ we need to set the value of $\epsilon_{k,h}$ appearing in condition~(\ref{eq:epk}). Since the PDF of $\bu_k(\infty)$ is Gaussian, a meaningful index of its variability is the standard deviation. Thus, a reasonable choice is $\epsilon_{k,h}=10^{-1} \sqrt{\V_h [\bu_k(\infty)]}$, which is used in all the experiments.

It is worth stressing that the distribution of the steady-state variable $\by_k(\infty)$ is, in general, far from being a Gaussian-shaped function. Due to the structure of Eq.~(\ref{eq:finapp}), the CDF of $\by_k(\infty)$ exhibits Gaussian-like increasing regions interleaved by regions where the CDF is constant. 
When $a_k$ gets large, however, the relative weight of the discrete component $\bz_k(\infty)$ decreases, and the distribution of the $\by_k(\infty)$ tends to approach the Gaussian distribution of the continuous component $\bu_k(\infty)$.
This is in agreement with the expected impact of the system parameters, as discussed in Sec.~\ref{sec:DSSS}: for larger $a_k$ and/or $|\N_k|$ a smoother CDF is obtained.
The curves shown in Figs.~\ref{fig:gausH0} and~\ref{fig:gausH1} confirm this behavior.

\subsection{Exponential Observations} \label{sec:expo}

Let $\br \sim \cE(\lambda)$ denote an exponential random variable of parameter $\lambda$, with PDF
$f(r)=\lambda \exp(-\lambda r)$, for $r \ge 0$.
Suppose that the i.i.d.\ observations at the various nodes are so distributed: $\br \sim \cE(\lambda_0)$ under $\cH_0$, and $\br \sim \cE(\lambda_1)$ under~$\cH_1$, with $0<\lambda_1<\lambda_0$. 
Assume also that the marginal statistics of the nodes are the log-likelihood ratios and that $\gamma_{\rm loc}=0$, see~(\ref{eq:logli}) and~(\ref{eq:q}). Straightforward algebra shows that $\bx =(\lambda_0-\lambda_1) \br  - \log (\lambda_0/\lambda_1)$. 

\begin{figure}
\centering 
\includegraphics[width =200pt]{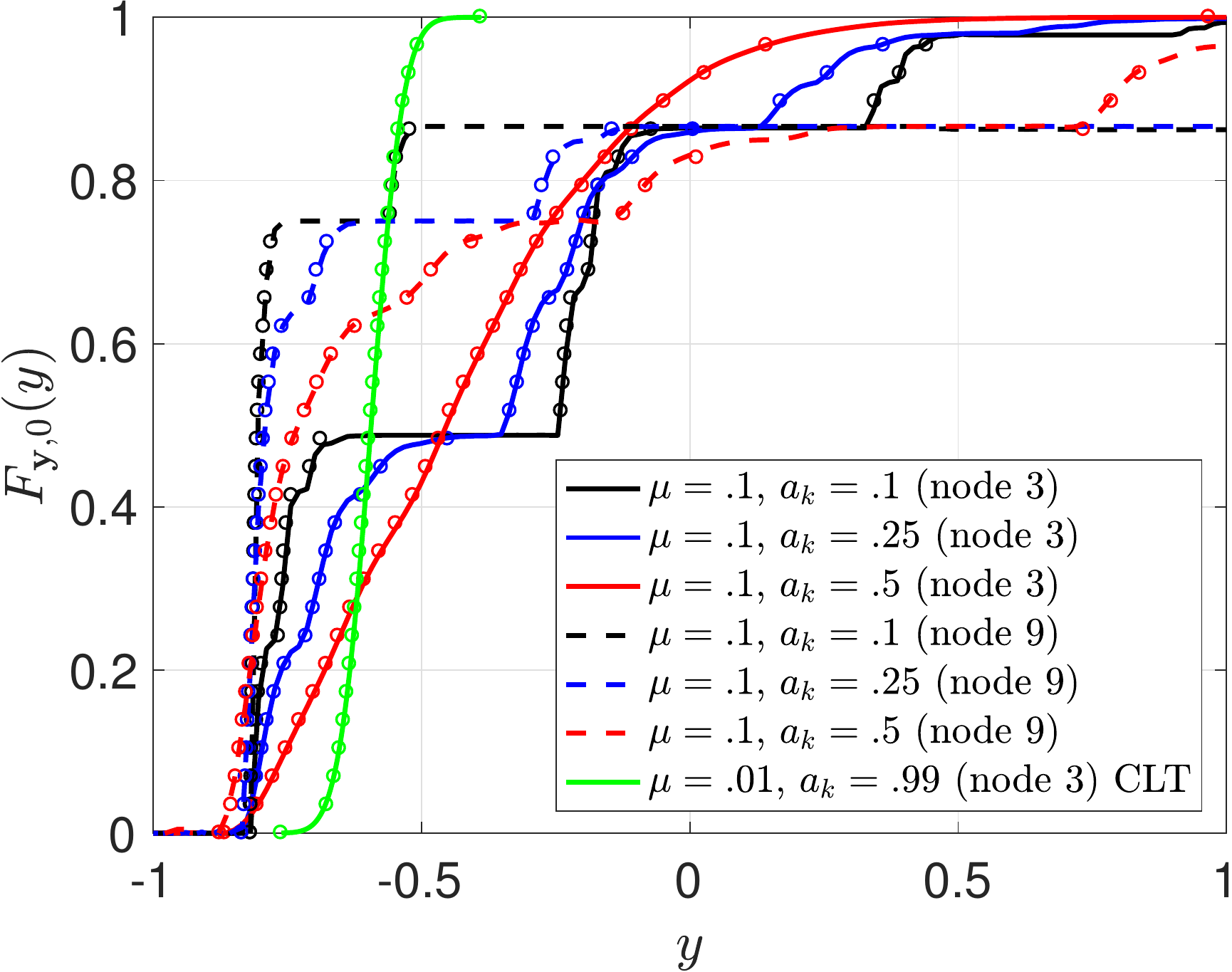}
  \caption{Exponential example under hypothesis $\cH_0$, with $\lambda_e=5$. It is shown the CDF $F_{\by,0}(y)$ of the steady-state variable $\by_k(\infty)$ for node $k=3$ (solid curves) and $k=9$ (dashed), obtained by the approximations developed in this paper. The symbols ``o'' show the results of computer experiments.}
      \label{fig:expoH0}
 \end{figure}

\begin{figure}
\centering 
\includegraphics[width =200pt]{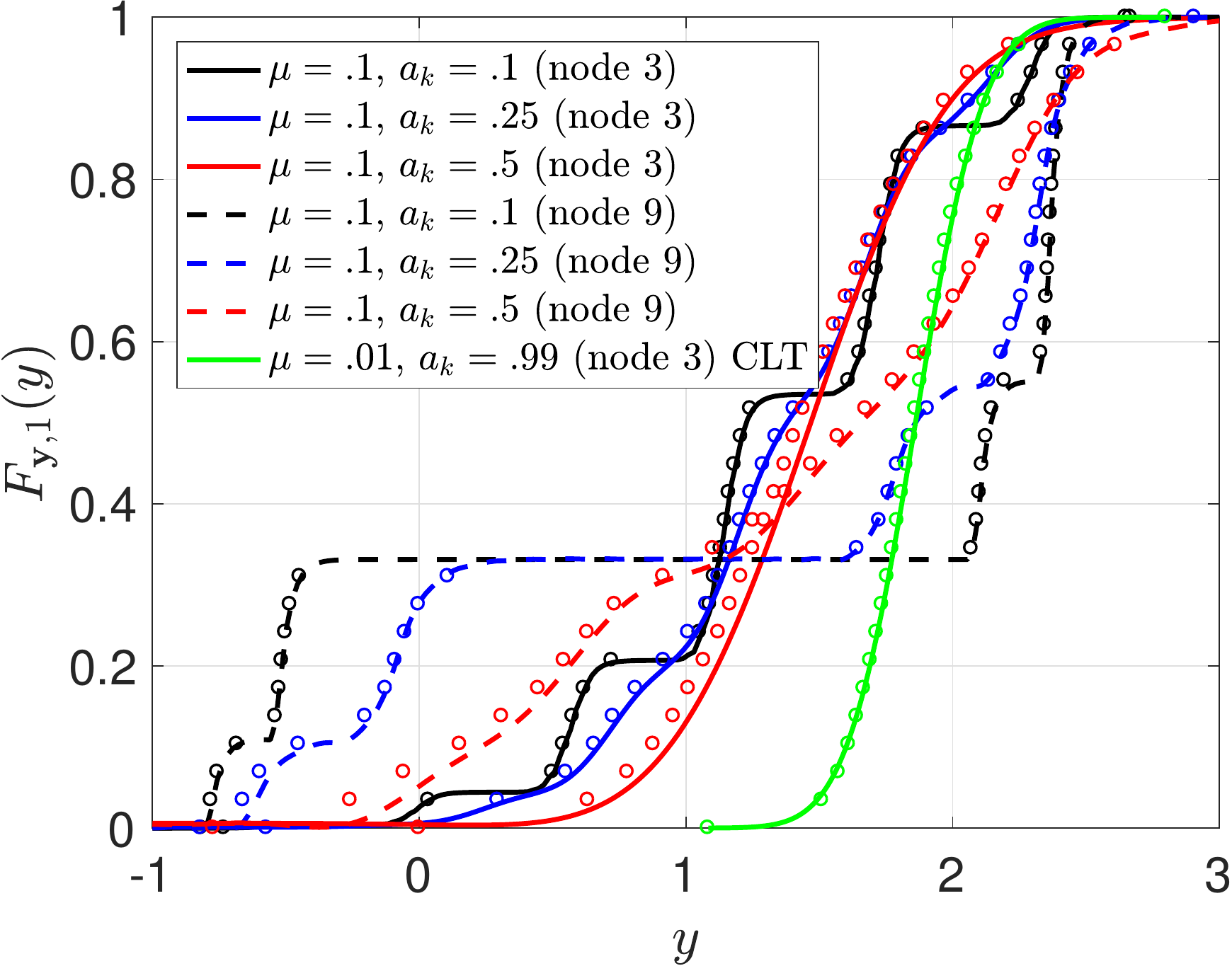}
  \caption{Exponential example under hypothesis $\cH_1$, with $\lambda_e=5$. It is shown the CDF $F_{\by,1}(y)$ of the steady-state variable $\by_k(\infty)$ for node $k=3$ (solid curves) and $k=9$ (dashed), obtained by the approximations developed in this paper. The symbols ``o'' show the results of computer experiments.}
      \label{fig:expoH1}
 \end{figure}

It is convenient to introduce the normalized parameter $\lambda_e \dfz \lambda_0/\lambda_1>1$.
Under $\cH_0$, 
the PDF of $\bx$ is 
\beq
f_{\bx,0}(x) = \left \{
\begin{array}{ll}
\frac{\lambda_e}{\lambda_e-1} e^{-\frac{\lambda_e}{\lambda_e-1} \left (x+\log \lambda_e \right )}, & x \ge -\log \lambda_e,  \\
0, & \textnormal{otherwise} .
\end{array}
\right .
\eeq
The mean and the variance of $\bx$ are $\E_0 \bx = 1-  \lambda_e^{-1} -\log \lambda_e$, and $\V_0 \bx  = \left ( 1-  \lambda_e^{-1} \right )^2$, while the 
log-characteristic function, see~(\ref{eq:logcf}), is
\beq
\Phi_{\bx,0}(t) = \log \frac{\lambda_e^{1-j \, t}}{\lambda_e - j \, t  \, (\lambda_e-1)}.
\label{eq:Phixexpo0}
\eeq
This function admits the series expansion $\Phi_{\bx,0}(t) = \sum_{i=1}^\infty \varphi_{n,0} t^n$, with coefficients
\beqa
\varphi_{n,0}= \left \{
\begin{array}{ll}
\left ( 1- \lambda_e^{-1} -\log \lambda_e \right ) j  ,& n=1, \\
\frac 1 n \left [\left ( 1- \lambda_e^{-1}\right ) j \, \right ]^n , & n>1 ,
\end{array}
\right .
\label{eq:cn0expo}
\eeqa
and radius of convergence $\tau_{\bx,0}=\frac{\lambda_e}{\lambda_e-1}$.

Using~(\ref{eq:Phixexpo0})-(\ref{eq:cn0expo}) in~(\ref{eq:appnew2}) gives the CDF $F_{\bu,0}(u)$, $u \in \Re$, of the continuous component $\bu_k(\infty)$, which according to~(\ref{eq:finapp}) can be used in combination with the second-order approximation developed for $\bz_k(\infty)$ in Sec.~\ref{sec:z}, to yield the CDF $F_{\by,0}(y)$, $y \in \Re$, of the steady-state variable $\by_k(\infty)$. 

Actually, to get the approximate CDF  $F_{\bu,0}(u)$ and the final $F_{\by,0}(y)$ we need to set the values of $\delta$, $\bar n$, $\bar m$ [see~(\ref{eq:appnew2})], and $\epsilon_{k,h}$ [see condition~(\ref{eq:epk})]. 
Clearly, $\bar n=\lfloor \frac{\tau_{\bx,0}}{\eta_k  \delta}-1 \rfloor$ from~(\ref{eq:maxnew}), while the choice of $\delta$ and $\bar m$ is discussed in Appendix~\ref{app:delta}.
As to $\epsilon_{k,h}$, since the PDF of $\bu_k(\infty)$ is unimodal and smooth, a reasonable choice is $\epsilon_{k,h}=10^{-1} \sqrt{\V_h [\bu_k(\infty)]}$, as done in the Gaussian case.

Consider now hypothesis $\cH_1$. In this case 
the PDF of $\bx$ is 
\beq
f_{\bx,1}(x) = \left \{
\begin{array}{ll}
\frac{1}{\lambda_e-1} e^{-\frac{1}{\lambda_e-1} \left (x+\log\lambda_e \right )}, & x \ge -\log \lambda_e,  \\
0, & \textnormal{otherwise} .
\end{array}
\right .
\label{eq:fx1}
\eeq
The mean and the variance of this random variable are $\E_1 \bx = \lambda_e -1 -\log  \lambda_e$, and $\V_1 \bx  = 5- 2\, \lambda_e + \lambda_e^2$.
The log-characteristic function corresponding to PDF~(\ref{eq:fx1}) can be easily computed, yielding:\footnote{Alternatively, a known property of the log-likelihood ratio can be exploited~\cite[Eq.\ (90), p. 44]{vantrees-book1}: For $x \in \Re$, let $f_{\bx,h}(x)$ be the PDF of the log-likelihood ratio~$\bx$. Then, one gets $f_{\bx,1}(x)=f_{\bx,0}(x)e^x$, which in terms of log-characteristic functions becomes $\Phi_{\bx,1}(t) = \Phi_{\bx,0}(t-j)$.}
\beq
\Phi_{\bx,1}(t) = \log \frac{\lambda_e^{-j \, t}}{1 - j \, t \, (\lambda_e-1)},
\label{eq:Phixexpo1}
\eeq
which again admits a series expansion in the form $\Phi_{\bx,1}(t) = \sum_{i=1}^\infty \varphi_{n,1} t^n$, with coefficients
\beqa
\varphi_{n,1}= \left \{
\begin{array}{ll}
\left ( \lambda_e -1-\log \lambda_e\right ) j  ,& n=1, \\
\frac 1 n \left [\left ( \lambda_e -1 \right ) j \, \right ]^n , & n>1 ,
\end{array}
\right .
\eeqa
and radius of convergence $\tau_{\bx,1}=\frac{1}{\lambda_e-1}$.
As before, exploiting these formulas in Theorem 1 gives the desired CDF $F_{\bu,1}(u)$ of the steady-state  continuous component $\bu_k(\infty)$, which combined as shown in~(\ref{eq:finapp}) with the approximation developed for $\bz_k(\infty)$, yields the CDF $F_{\by,1}(y)$.
The quantities $\delta$, $\bar n$, $\bar m$, and $\epsilon_{k,h}$, are computed as under $\cH_0$.

\begin{figure}
\centering 
\includegraphics[width =190pt]{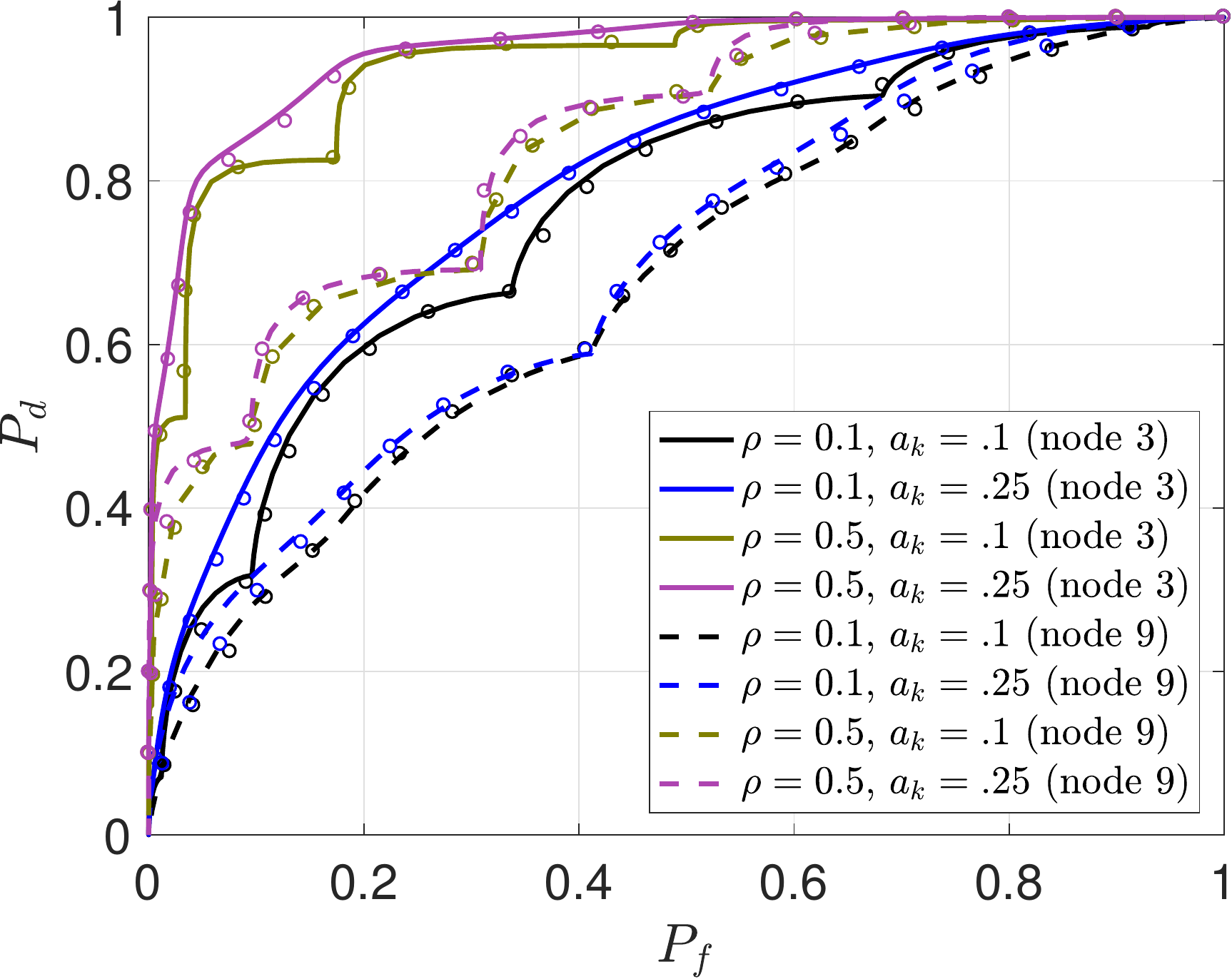}
  \caption{ROCs of agents $k=3$ and $k=9$ for the Gaussian example of Sec.\ref{sec:gau}, with $ \mu=0.1$ and different combinations of $\rho$ and $a_k$.
  Solid and dashed curves refer to the theoretical expressions derived in Sec.~\ref{sec:SSA} while the small circles denote simulation points.}
      \label{fig:rocg}
 \end{figure}

The theoretical CDFs $F_{\by,0}(y)$ and $F_{\by,1}(y)$ are compared to the results of Monte Carlo experiments in Figs.~\ref{fig:expoH0} and~\ref{fig:expoH1}. 
Figure~\ref{fig:expoH0} shows the CDF of $\by_k(\infty)$ under $\cH_0$, for the highly-connected node $k=3$ with $5$ neighbors, and for the weakly connected node $k=9$ with only one neighbor. In these experiments we set $\lambda_e=5$, yielding $(1-p_f)\approx 0.87$, 
$\mu=0.1$, and consider three different values of $a_k=0.1, 0.25, 0.5$. The curves represent the theoretical CDFs, with solid lines for $k=3$, and dashed lines for $k=9$. The symbol ``o'' denotes values obtained by computer experiments. 
The prescriptions of Theorem~2 are confirmed by the curve in green, which refers to $\mu=0.01$ and $a_k=0.99$, and by the correspondent simulation points obtained with $n=1000$ iterations. The matching between theory and simulation points is practically perfect.

The same comparison between theoretical predictions and results of computer experiments is conducted under $\cH_1$, in which case $p_d \approx 0.67$. The results are reported in Fig.~\ref{fig:expoH1}. Again, we see that the matching between theory and simulation is very satisfying.
We also note that the effect of the network topology, encoded in $\N_k$, and the system parameters $a_k$ and~$\mu$, is as predicted, see discussion in Sec.~\ref{sec:DSSS}.

\subsection{Detection Performance} \label{sec:detp}

Consider the statistical test~(\ref{eq:test}) at steady-state. We define the system-level false alarm probability of node $k$ as $P_f \dfz \P_0(\by_k(\infty)>\gamma)=1-F_{\by,0}(\gamma)$, and the system-level detection probability of node $k$ as $P_d \dfz \P_1(\by_k(\infty)>\gamma)=1-F_{\by,1}(\gamma)$. The quantities $P_f$ and $P_d$ depend on the node index $k$ and on the threshold $\gamma \in \Re$, but for notational convenience these dependencies are not made explicit. 
By computing $F_{\by,0}(\gamma)$ for a wide range of values of $\gamma$, the system designer may choose the threshold value, say it $\gamma^\ast$, such that a prescribed value of false alarm $P_f^\ast$ is achieved. Then, the corresponding value of detection probability $P_d^\ast=1-F_{\by,1}(\gamma^\ast)$ can be computed.
The function relating $P_d$ to $P_f$ can be obtained by varying $\gamma^\ast$, and is usually referred to as the ROC (receiver operating characteristic)~\cite{vantrees-book1}. 
We see that the ROCs of the diffusion network under one-bit messaging are easily derived by exploiting the theoretical CDFs $F_{\by,h}(y)$, $h=0,1$, obtained in this paper. The effect on the ROCs of the system parameters --- node connectivity~$|\N_k|$, self-combination coefficient~$a_k$, and step-size~$\mu$ --- can be inferred by the effect of these parameters on the CDFs, which was discussed in the previous sections.

\begin{figure}
\centering 
\includegraphics[width =190pt]{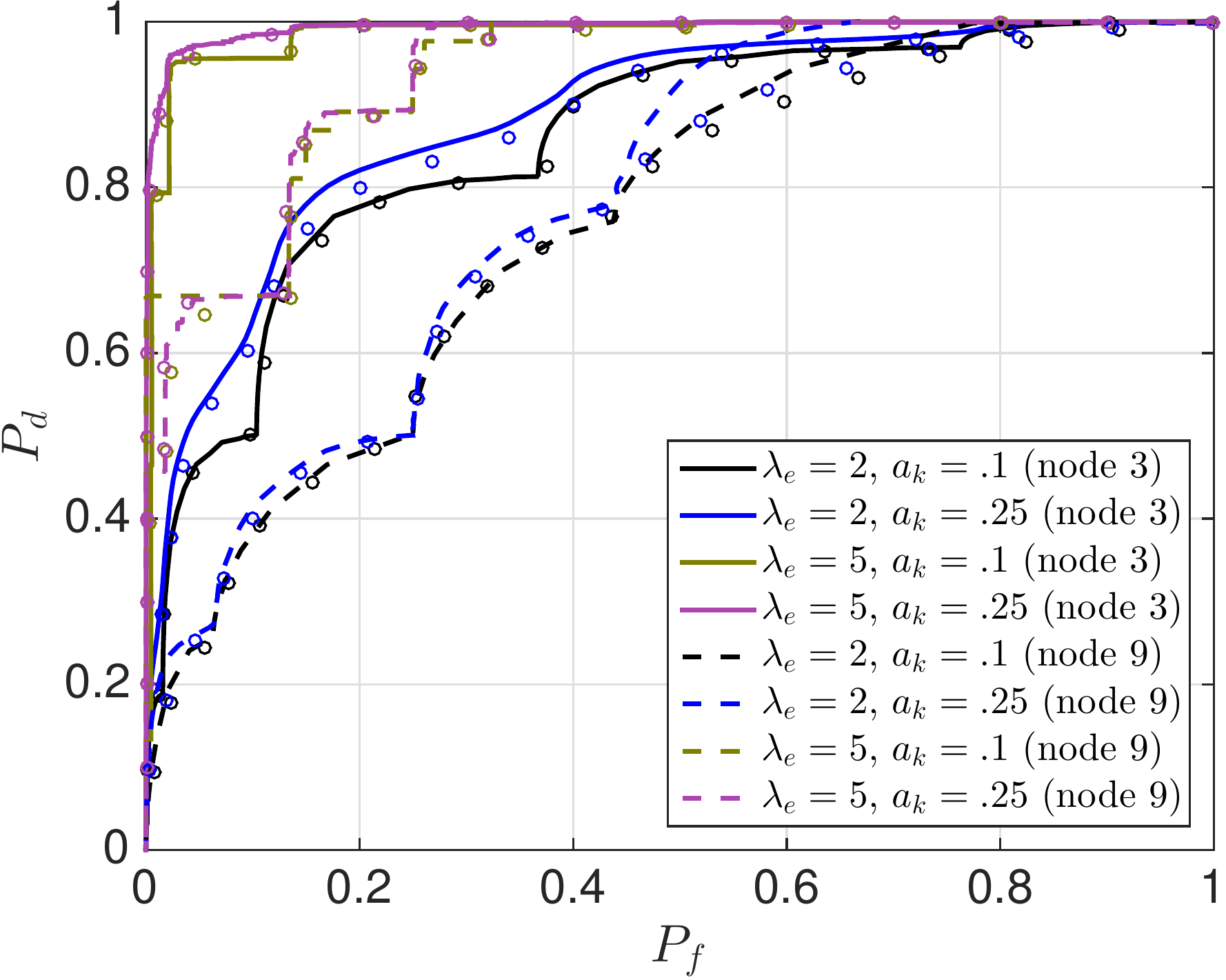}
  \caption{ROCs of agents $k=3$ and $k=9$ for the exponential example of Sec.~\ref{sec:expo}, with $ \mu=0.1$ and different combinations of $\lambda_e$ and $a_k$. Solid and dashed curves refer to the theoretical expressions derived in Sec.~\ref{sec:SSA} while the small circles denote simulation points.}
      \label{fig:roce}
 \end{figure}

Throughout this section the step size $\mu$ is maintained constant to the typical value $0.1$, see~\cite{BracaetalIT,MattaSIPN16}.
Figures~\ref{fig:rocg} and~\ref{fig:roce} show the system-level ROC of node $k$ for the Gaussian and for the exponential cases, respectively.
Consider first the Gaussian case in Fig.~\ref{fig:rocg}, and let us focus on node $k=3$ (solid curves) and $\rho=0.1$. We see that the ROC curve is rather nontrivial  when $a_k=0.1$, and tends to be smoother, closer to a concave curve, and closer to the ideal curve ($P_d=1$ for all $P_f>0$) for $a_k=0.25$. 
The irregular non-concave behavior\footnote{A concave ROC can be obtained by randomization, see~\cite{poorbook}.} of the ROC is an immediate consequence of the irregular behavior of the CDFs $F_{\by,h}(y)$, $h=0,1$, namely of the highly multimodal densities arising from the interaction between the continuous and the discrete components.
For small values of $a_k$ much credit is given to information coming from the neighbor nodes. Since the information coming from the neighbor agents is in the form of \emph{marginal} decisions, the more irregular shape of the ROCs can be explained by the stronger effect of the discrete component with respect to the continuous one.
When~$a_k$ grows the distributions of the steady-state $\by_k(\infty)$ under the two hypotheses tend to be more separated and indeed the performance improves. Recall that the smaller~$a_k$ is, the better the adaptive properties of the system become. Thus, the improved detection performance obtained when $a_k$ grows is paid in terms of a slower adaptivity to changes in the state of nature, which is a manifestation of an inherent system tradeoff.
The same behavior is observed for $\rho=0.5$. Clearly, the larger~$\rho$ is, the larger the detection performance becomes and indeed the ROC curves for $\rho=0.5$ stand above the curves for $\rho=0.1$.

Consider next the ROC curves of the less-connected node $k=9$ (dashed lines). These curves show the same qualitative behavior as node $3$. However, the lower connectivity of node~$9$ is paid in terms of a worse detection performance, which shows the effect of the network topology and the benefit of a stronger system connectivity.

It is worth noting from Fig.~\ref{fig:rocg} that the simulation points closely fit the theoretical curves, and some discrepancy can be only observed for node $k=9$ at small $\rho$ and large $a_k$. In general, the accuracy of the theoretical predictions improves with a larger number of agent neighbors, larger values of $\rho$, and smaller values of~$a_k$.

The same analysis conducted for the Gaussian example is repeated in Fig.~\ref{fig:roce} for the exponential example of Sec.~\ref{sec:expo}, and similar considerations apply.

\section{Extensions}
\label{sec:added}
In this section we briefly discuss two possible extensions of our work.
First, in many practical scenarios, especially when the network covers wide geographical areas, the assumption of identical distribution of the data collected by the remote agents
may not be valid. The diffusion rule~(\ref{eq:new1})-(\ref{eq:new2}), however, remains unaltered because the one-bit messages received by agent $k$ are binary flags denoting the local decisions of its neighbors. These flags can be incorporated in~(\ref{eq:new2}) as $\E_1 \bx$ or $\E_0 \bx$, where the expectations are taken with respect to the statistical distributions of agent $k$'s observations, regardless of the distributions characterizing the observations of the agents that originate the messages. One difference with respect to the analysis developed in the previous sections is that the marginal detection and false alarm probabilities, see~(\ref{eq:pdpf}), can differ from agent to agent. Thus, the characterization of the discrete component in Sec.~\ref{sec:z} must be revisited in order to take into account the non-identical distributions of the normalized binary variables $\bb_\ell(i)$ appearing in~(\ref{eq:bb}).

Another case of practical relevance is that of multihypothesis tests, in which the task of the network is to decide among~$H>2$ possible states of nature $\cH_0, \dots, \cH_{H-1}$. To see how the tools developed in this article can be exploited in this scenario, recall from~(\ref{eq:q}) that the messages exchanged among agents can be interpreted as the agents' marginal decisions. Accordingly, let us assume that 
there are~$H$ marginal decision statistics $\bx_{k}^{(h)}(n)=\log f_{\br,h}(\br_k(n))$, $h=0,\dots,H-1$, in place of the single variable shown in~(\ref{eq:logli}), 
and the message delivered by agent~$\ell$ at time~$n$ consists of its local decision $\kappa_\ell(n)=\arg\max_{h=0,\dots,H-1}\ \bx_{k}^{(h)}(n)$, with $\kappa_\ell(n)\in \{0,\dots,H-1\}$. Then, $H$ local state variables $\by_k^{(h)}(n)$, $h=0,\dots,H-1$, can be defined, and we have the following 
straightforward extension of~(\ref{eq:new1})-(\ref{eq:new2}):
\begin{subequations}
\begin{align}
&\bv_k^{(h)}(n)=\by_k^{(h)}(n-1)+\mu [\bx_k^{(h)}(n)-\by_k^{(h)}(n-1)], \label{eq:mh1}\\
&\by_k^{(h)}(n)= a_{kk} \bv_k^{(h)}(n)  + \sum_{\ell\neq k} a_{k \ell} \, \E_{\kappa_\ell(n)} \bx^{(h)}, \, \, \, \, \, n \ge 1, \label{eq:mh2}
\end{align}
\end{subequations}
for  $h=0,\dots,H-1$.
The decision of agent $k$ at steady-state is $\arg\max_{h=0,\dots,H-1}\ \by_{k}^{(h)}(\infty)$. 
Note that in the diffusion scheme~(\ref{eq:mh1})-(\ref{eq:mh2}), each of the $H$ local state variables $\{\by_k^{(h)}(n)\}$
evolves exactly  as does the single variable $\by_k(n)$ dealt with in the previous sections, and the analysis is similar.
In particular, the analysis of the continuous component closely follows the one developed in this article, while the analysis of the discrete component requires further elaboration to generalize the methods of Sec.~\ref{sec:z} to the presence of $H>2$ local decisions. The details are left for future investigations.

\vspace*{-1pt}
\section{Summary \& Future Work}
\label{sec:concl}

Inter-agent communication constraints is an important aspect of multi-agent systems for on-line adaptation and learning. 
In this work we consider a suitably modified version of the classical ATC diffusion rule, under the constraint that the inter-agent links support only one-bit messages.
For the proposed one-bit messaging network, a careful investigation of the steady-state distribution of the agents' state is carried on, and the resulting performance curves reveal sophisticated/nontrivial behavior which is predicted well by the derived theoretical expressions. The role played by the main system parameters --- step-size $\mu$, self-combination coefficients $\{a_k\}$, and network topology $\N_k$ --- is investigated.

The results provided in this paper can be exploited for the design and analysis of more complex decision systems, some of which are mentioned in Sect.~\ref{sec:added}. Further examples of these possible future extensions include the analysis of networks where the messages can be delivered to farther agents by multi-hop communications, managing the tradeoff between the number of delivered bits per link usage and the number of channel accesses, and addressing more advanced hypothesis testing problems (composite hypotheses, sequential detection, robust techniques). Incorporating security issues is another important direction for future studies such as addressing the presence of adversary agents (e.g., Byzantine sensors~\cite{marmatton-byzantine}) that conspire for impairing the network by delivering malicious messages.

\vspace*{-3pt}
\begin{appendices}
\numberwithin{equation}{section}

\section{Proof of Theorem 1}
\label{app:newappth1}

Consider Eq.~(\ref{eq:winf}). 
Before letting $n \rightarrow \infty$, the following equality in distribution holds
\beq
 \sum_{i=1}^{n}  \eta_k^{i-1} \, \bx_k(n-i+1) \deq \sum_{i=1}^{n}  \eta_k^{i-1} \, \bx_k(i).
 \label{eq:ind}
\eeq
Recalling from~(\ref{eq:eta}) that $\eta_k \in (0,1)$, we have
$\E_h[\sum_{i=1}^{n}  \eta_k^{i-1} \, \bx_k(i)]$ $\rightarrow$ $\E_h \bx /(1-\eta_k)$, for $n \rightarrow \infty$. Thus, if the sum $\sum_{i=1}^{n}  \eta_k^{i-1} \, (\bx_k(i)- \E_h \bx)$ of zero-mean independent variables  converges with probability one, so does the right-hand side of~(\ref{eq:ind}) for the linearity of the almost everywhere limit~\cite{shao}. 

The sum of zero-mean independent random variables converges 
either with probability one or with probability 
zero 
and, in fact, it converges with probability one if the sum of the variances converges~\cite[Th. 22.6]{billingsley-book2}. In our case $\lim_{n \rightarrow \infty} \sum_{i=1}^n \V_h [\eta_k^{i-1} \,( \bx_k(i)-\E_h \bx)] = \V_h \bx /(1-\eta_k^2) < \infty$, and this proves the convergence with probability one of $\sum_{i=1}^{n}  \eta_k^{i-1} \, \bx_k(i)$ to the random variable $\bw_k^\star$, defined in~(\ref{eq:winf}).
Convergence with probability one implies convergence in distribution~\cite{shao} and, again from~(\ref{eq:winf}), 
this shows that $\sum_{i=1}^{n}  \eta_k^{i-1} \, \bx_k(n-i+1)$, hence $\bu_k(n)$, converges in distribution.

To prove~(\ref{eq:Phiw}), suppose that~(\ref{eq:serieshyp}) holds. The series on the right-hand side of~(\ref{eq:serieshyp}) converges absolutely and uniformly for $|t|<\tau_{\bx,h}$~\cite[Th. 93]{knopp}~\cite[10.5, p. 198]{Rudin}. A necessary and sufficient condition for the absolute and uniform convergence in the interval $|t|<\tau_{\bx,h}$ of this series, is
that its coefficients $\{\varphi_{n,h}\}$ satisfy $\limsup_{n\rightarrow \infty} |\varphi_{n,h}|^{\frac 1 n}=\tau_{\bx,h}^{-1}$, see e.g.,~\cite[Th. 94]{knopp}. 
The quantity $\tau_{\bx,h}^{-1}$ is interpreted as $0$ when $\tau_{\bx,h}=\infty$.
For all $0 < \eta_k < 1$ we have $\lim_{n \rightarrow \infty} (1-\eta_k^n)^{- \frac 1 n }=1$, yielding
$\limsup_{n \rightarrow \infty} \left |\frac{\varphi_{n,h}}{1-\eta_k^n} \right |^{\frac 1 n} = \tau_{\bx,h}^{-1}$.
Thus, the radius of convergence of the series 
\beq
\sum_{n=1}^\infty \frac{\varphi_{n,h}}{1-\eta_k^n} t^n
\label{eq:series2}
\eeq
is $\tau_{\bx,h}$, namely, (\ref{eq:series2})~converges absolutely and uniformly for $|t| < \tau_{\bx,h}$.

To show that the series~(\ref{eq:series2}) represents the log-characteristic function of $\bw_k^\star$, note that, from~(\ref{eq:key2}),
\beqa
\Phi_{\bw,h}(t) &=& \log \E_h  e^{j t \bw_k^\star}=\log \E_h  e^{j t (\eta_k \bw_k^\star+\bx)} \nonumber \\
 &=& \log \E_h  e^{j t \eta_k \bw_k^\star} + \log \E_h  e^{j t \bx} .
\eeqa
Therefore, for all $t \in \Re$, the following linear functional equation in the unknown $\Phi_{\bw,h}(t)$ holds:
\beq
\Phi_{\bw,h}(t)-\Phi_{\bw,h}(\eta_k\,  t)=\Phi_{\bx,h}(t). 
\label{eq:funeq}
\eeq
We now substitute the function $\Phi_{\bw,h}(t)$ appearing in~(\ref{eq:funeq}) by the power series~(\ref{eq:series2}). 
Before, note that:
\beq
\lim_{n \rightarrow \infty} \left |\frac{\eta_k^n}{1-\eta_k^n} \right |^{\frac 1 n}\hspace*{-5pt}=\eta_k \;\; \Rightarrow \; \;\limsup_{n \rightarrow \infty} \left |\frac{\eta_k^n\varphi_{n,h}}{1-\eta_k^n} \right |^{\frac 1 n}\hspace*{-5pt}=\frac{\eta_k}{\tau_{\bx,h}}, 
\label{eq:alimit}
\eeq
which shows that the radius of convergence of $\sum_{n=1}^\infty \frac{\varphi_{n,h}}{1-\eta_k^n} \eta_k^n t^n$ is $\frac{\tau_{\bx,h}}{\eta_k}>\tau_{\bx,h}$.
Since absolutely convergent power series can be added term by term~\cite[Th. 3, p.\ 134]{knopp}, using~(\ref{eq:series2}) in place of $\Phi_{\bw,h}(t)$, from~(\ref{eq:funeq}) one gets:
\begin{align}
&\sum_{n=1}^\infty \frac{\varphi_{n,h}}{1-\eta_k^n} t^n - \sum_{n=1}^\infty \frac{\varphi_{n,h}}{1-\eta_k^n}\eta_k^n  t^n 
=\sum_{n=1}^\infty \frac{\varphi_{n,h}}{1-\eta_k^n} (1-\eta_k^n)t^n \nonumber \\
&\qquad = \sum_{n=1}^\infty \varphi_{n,h}t^n = \Phi_{\bx,h}(t), \qquad |t| < \tau_{\bx,h},
\end{align}
which shows that~(\ref{eq:funeq}) is verified in the interval $|t| < \tau_{\bx,h}$, thus proving~(\ref{eq:Phiw}).

The solution $\Phi_{\bw,h}(t)=\sum_{n=1}^\infty \frac{\varphi_{n,h}}{1-\eta_k^n} t^n$ to~(\ref{eq:funeq}) is unique among the class of functions that admit an absolutely convergent power series in a neighbor of the origin $|t| < t_0$, for some $t_0>0$. Indeed, suppose that $g(t)=\sum_{n=1}^\infty d_{n} t^n$ is another solution in that class. Then, $g(\eta_k \, t)=\sum_{n=1}^\infty d_{n} \eta_k^n t^n$ is absolutely convergent for $|t| < t_0/\eta_k$. We can add term by term these series, obtaining $g(t)-g(\eta_k \, t)=\sum_{n=1}^\infty d_{n} (1-\eta_k^n)t^n=\sum_{n=1}^\infty \varphi_{n,h} t^n$, where the last equality holds because, by assumption, $g(t)$ solves~(\ref{eq:funeq}) in $|t|<t_0$. The uniqueness of the power series expansion~\cite[Th. 97, p.\ 172]{knopp} implies $d_{n}=\varphi_{n,h}/(1-\eta_k^n)$, for all $n$. 
Because the coefficients are the same, we conclude that $g(t)=\Phi_{\bw,h}(t)$ for $|t|<\tau_{\bx,h}$.

For $\tau_{\bx,h}=\infty$, Eq.~(\ref{eq:newth}) immediately follows by a known formula for computing the CDF of $\bw_k^\star$, assumed everywhere continuous, from its characteristic function $\exp[\Phi_{\bw,h}(t)]$\cite[Eq.~(2)]{shephard,gurland1948}, upon recalling from~(\ref{eq:uinf}) that the distribution of $\bu_k(\infty)$ is simply a scaled version of the distribution of $\bw_k^\star$.

To arrive at~(\ref{eq:appnew}), let $0<\tau_{\bx,h}\le\infty$. From~\cite[Eq.~(7)]{shephard,davies} we have a formula for obtaining the CDF $F_{\bw,h}(u)$, $u \in \Re$, of $\bw_k^\star$ from the corresponding log-characteristic function $\Phi_{\bw,h}(t)$, $t \in \Re$. Expressed in terms of $F_{\bu,h}(u)=F_{\bw,h}\big(\frac{u}{\mu a_k}\big)$,
the formula is:
\begin{align}
F_{\bu,h}(u) = \frac 1 2 - \frac {2}{\pi} \sum_{n=0}^\infty \Omega_n (u,\delta )
 + \chi_{\bu}(u,\delta),
\label{eq:bigappnew}
\end{align}%
where $\Omega_n(u,\delta)$ is given in~(\ref{eq:notation}), and the ``error'' term verifies:
\beq
\small
|\chi_{\bu}(u,\delta)| \le \max \Big \{ \hspace*{-2pt}  F_{\bu,h}\Big(u - \frac{2 \pi a_k \mu}{\delta} \Big ) ,  1- F_{\bw,h}\Big(u - \frac{2 \pi a_k \mu}{\delta}\Big ) \Big \} .
\label{eq:uppboundnew}
\eeq
Fix $\epsilon^\prime>0$, and suppose that $\delta$ is such that the second condition in~(\ref{eq:maxnew}) is verified. 
Then, from~(\ref{eq:bigappnew})-(\ref{eq:uppboundnew}) we see that
the absolute value of the error incurred when computing $F_{\bu,h}(u)$ by the expression 
$\frac 1 2 - \frac {2}{\pi} \sum_{n=0}^\infty \Omega_n(u,\delta)$ is $\le \epsilon^\prime/2$.
Next, suppose that $\delta$ also verifies the first condition in~(\ref{eq:maxnew}), namely truncating the series implies an error not larger than $\epsilon^\prime/2$. 
Then:
\begin{align}
F_{\bu,h}(u) = \frac 1 2 - \frac {2}{\pi} \sum_{n=0}^{\bar n} \Omega_n (u,\delta )
 + \Delta_{\bu}(u,\delta), \label{eq:clean}
\end{align}
where we have denoted $\lfloor \frac{\tau_{\bx,h}}{\eta_k  \delta}-1 \rfloor$ by $\bar n$. In~(\ref{eq:clean}), by triangular inequality, we have $|\Delta_{\bu}(u,\delta)|<\epsilon^\prime$.

Consider now the function $\Phi_{\bw,h}(\cdot)$ appearing at the exponent of~(\ref{eq:notation}). We have: 
\begin{subequations} \begin{align}
\Phi_{\bw,h}(t)&=\Phi_{\bx,h}(t)+\Phi_{\bw,h}(\eta_k \, t) \label{subeq:start} \\
&= \Phi_{\bx,h}(t)+\sum_{m=1}^{\infty} \frac{\varphi_{m,h}}{1-\eta_k^m} \eta_k^m t^m, \label{subeq:series}
\end{align} \end{subequations}
where~(\ref{subeq:start}) is the same of~(\ref{eq:funeq}), and the series expansion in~(\ref{subeq:series}) has radius of convergence $\tau_{\bx,h}/\eta_k$
because of~(\ref{eq:alimit}). Noting that $\Omega_n(u,\delta)$ in~(\ref{eq:notation}) involves $\Phi_{\bw,h}((2n+1)\delta/2)$, whose argument lies in the region of convergence of expansion~(\ref{subeq:series}) for all $n \le \lfloor \frac{\tau_{\bx,h}}{\eta_k  \delta}-1 \rfloor$, we can use~(\ref{subeq:series}) in the expression of $\Omega_n(u,\delta)$ appearing in~(\ref{eq:clean}), yielding~(\ref{eq:appnew}), which proves claim $iii)$.

\section{Proof of Theorem 2}
\label{app:theo2}

The material in this section is adapted from~\cite[Theorems 27.2 and 27.3]{billingsley-book2}. 
From~(\ref{eq:rec2}), let us denote by 
\beqa
\bx_k^\prime(n)&=&\bx_k(n)-\E_h \bx, \label{eq:note1} \\
\widetilde \bx_k^\prime(n)&=&\widetilde \bx_k(n)-\E_h \widetilde \bx, \label{eq:note2}
\eeqa the zero-mean versions of $\bx_k(n)$ and $\widetilde \bx_k(n)$, respectively, and
consider the normalized (zero-mean unit-variance) process
\beq
\bar \by_k(n)=\frac{\by_k(n)-\E_h[\by_k(n)]}{\sqrt{\V_h[\by_k(n)]}}=\sum_{i=1}^n \bar \bw_k (n,i),
\eeq 
where [see~(\ref{eq:array}),~(\ref{eq:dfzm}) and~(\ref{eq:dfzs})]:
\begin{align}
&\bar \bw_k(n,i)  \dfz   \frac{\bw_k(n,i)- \E_h [\bw_k(n,i)]}{s_k(n)}  \nonumber \\
& \quad = \frac{\eta_k^{i-1}}{s_k(n)} \left [ \mu a_k  \bx_k^\prime(n-i+1) 
 + \sum_{\ell \neq k} a_{k\ell}  \widetilde \bx_{\ell}^\prime(n-i+1)   \right ]. \label{eq:defw2}
\end{align} 
For $t \in \Re$,  let $\phi(t; n,i) \dfz \E_h  e^{jt \bar \bw_k(n,i) }$ 
be the characteristic function of $\bar \bw_k(n,i)$, where $j=\sqrt{-1}$. 
Denote by $\cI_C$ the indicator function of condition $C$. For sufficiently small $\epsilon>0$, and each fixed $t \in \Re$, one gets
\begin{align}
&\hspace*{-5pt}\left | \phi(t; n,i)  - \left (1 - \frac {t^2}{2}\V_h[\bar \bw_k(n,i)]\right ) \right | \label{eq:start} \\
&\hspace*{-5pt}    \le \E_h \left | e^{jt \bar \bw_k(n,i) } -\left (1+jt \bar \bw_k(n,i) - \frac{t^2}{2} \bar \bw_k^2(n,i)\right) \right |   \label{eq:li1} \\
&\hspace*{-5pt}   \le \E_h \left [\min\{|t \bar \bw_k(n,i)|^2,|t \bar \bw_k(n,i) |^3/6\}  \right ]\label{eq:li2} \\
&\hspace*{-5pt}   \le \E_h \left [ |t \bar \bw_k(n,i) |^3  \cI_ {|\bar \bw_k(n,i)| < \epsilon} \right ] /6 \nonumber  \\
& \quad  + \E_h \left [ |t \bar \bw_k(n,i) |^2  \cI_{|\bar \bw_k(n,i)| \ge  \epsilon} \right ] \label{eq:li3} \\
&\hspace*{-5pt}   \le \frac{1}{6}\epsilon |t|^3 \V_h[\bar \bw_k(n,i)]
 + t^2 \, \E_h \left [ \bar \bw_k^2(n,i)  \cI_{|\bar \bw_k(n,i)| \ge  \epsilon} \right ] \label{eq:li4}
\end{align}
where~(\ref{eq:li1}) follows by Jensens' inequality~\cite{CT2}; 
inequality~(\ref{eq:li2}) follows by the relationship 
\beq
\left | e^{jtz} -(1+jtz - \frac 1 2 t^2 z^2) \right | \le \min\{|tz|^2,|tz|^3/6\},
\label{eq:basmod}
\eeq
which holds for any $z \in \Re$, see, e.g.~\cite[26.4$_2$]{billingsley-book2}; inequality~(\ref{eq:li3}) is valid for any $\epsilon>0$;
finally, for $|\bar \bw_k(n,i)| < \epsilon$ we have $|\bar \bw_k(n,i) |^3 = \bar \bw_k^2(n,i)  |\bar \bw_k(n,i)| <  \bar \bw_k^2(n,i) \, \epsilon$, which gives~(\ref{eq:li4}).

Summing over $i=1,\dots,n$, inequality~(\ref{eq:start})-(\ref{eq:li4}) yields
\begin{align} 
&\sum_{i=1}^n \left | \phi(t;n,i)  - \left (1 - \frac {t^2}{2}\V_h[\bar \bw_k(n,i)]\right ) \right | \nonumber \\
& \le \frac 1 6 \epsilon |t|^3 + t^2 \sum_{i=1}^n \E_h \left [ \bar \bw_k^2(n,i)  \cI_{|\bar \bw_k(n,i)| \ge  \epsilon} \right ]  .
\label{eq:noyetinf}
\end{align}

Since $\epsilon>0$ can be made as small as desired, this means
\beq
\lim_{\eta_k\rightarrow 1} \lim_{n\rightarrow \infty} \sum_{i=1}^n \left | \phi(t;n,i)  - \left (1 - \frac {t^2}{2}\V_h[\bar \bw_k(n,i)]\right ) \right | 
=0,
\label{eq:hopeok}
\eeq
for all $t \in \Re$, provided that the \emph{Lindeberg condition}: $\forall \epsilon >0$, 
\beq
\lim_{\eta_k\rightarrow 1} \lim_{n\rightarrow \infty} \sum_{i=1}^n \E_h \left [ \bar \bw_k^2(n,i)  \cI_{|\bar \bw_k(n,i)| \ge  \epsilon} \right ]  =0,
\label{eq:linde}
\eeq
is verified.

Now, whatever $\epsilon$ is, for any $\beta>0$ 
\begin{align}
&\E_h \left [ \bar \bw_k^2(n,i)  \cI_{|\bar \bw_k(n,i)| \ge  \epsilon} \right ] \nonumber \\
& \; \le \E_h \left [ \bar \bw_k^2(n,i)  \left (\frac{|\bar \bw_k(n,i)|}{\epsilon }\right )^\beta  \cI_{| \bar \bw_k(n,i)| \ge  \epsilon}  \right ] 
\nonumber \\
& \; = \frac{1}{\epsilon^\beta }\, \E_h \left [|\bar \bw_k(n,i)|^{2 + \beta} \cI_{| \bar \bw_k(n,i)| \ge  \epsilon} \right ]  \nonumber \\
& \; \le \frac{1}{\epsilon^\beta }\, \E_h \left [|\bar \bw_k(n,i)|^{2 + \beta}\right ] .
\end{align}
This shows that Lindeberg condition~(\ref{eq:linde}) is implied by the following \emph{Lyapunov condition}: there exists $\beta>0$ such that 
\beq
\lim_{\eta_k \rightarrow 1}\lim_{n\rightarrow \infty}\sum_{i=1}^n \E_h \left [ |\bar \bw_k(n,i)|^{2 + \beta}   \right ] 
=0.
\label{eq:lyapo}
\eeq

Lemma 1 in Appendix~\ref{app:lyapo} shows that~(\ref{eq:lyapo}) holds true.
Therefore, we now proceed to conclude the proof of the theorem, assuming~(\ref{eq:lyapo}), which implies~(\ref{eq:linde}), which in turn implies~(\ref{eq:hopeok}).

The characteristic function of $\bar \by_k(n)$ is $\E_h e^{jt \sum_{i=1}^n \bar \bw_k(n,i)}$ $=\prod_{i=1}^n \phi(t;n,i)$, and we want to show that such characteristic function converges to the characteristic function $\E_h[e^{jt\,\bg}]= e^{-t^2/2}$ of a standard Gaussian random variable~$\bg$. To show this, 
note that $e^{-t^2/2}=\prod_{i=1}^n  e^{-\V_h [\bar \bw_k(n,i)]t^2/2}$, and consider the following:
\begin{align}
&\left | \E_h e^{jt \sum_{i=1}^n \bar \bw_k(n,i)} - e^{-t^2/2} \right | \nonumber \\
&\; = \left | \prod_{i=1}^n \phi(t;n,i) -  \prod_{i=1}^n  e^{-\V_h [\bar \bw_k(n,i)]t^2/2} \right | \nonumber \\
&\; =\left | \prod_{i=1}^n \phi(t;n,i) - \prod_{i=1}^n \left (1 - \frac {t^2}{2}\V_h[\bar \bw_k(n,i)]\right ) \right .\nonumber \\
& \quad \left . + \prod_{i=1}^n \left (1 - \frac {t^2}{2}\V_h[\bar \bw_k(n,i)]\right ) - \prod_{i=1}^n  e^{-\V_h [\bar \bw_k(n,i)]t^2/2} \right | \nonumber \\
& \; \le \left | \prod_{i=1}^n \phi(t;n,i) - \prod_{i=1}^n \left (1 - \frac {t^2}{2}\V_h[\bar \bw_k(n,i)]\right ) \right | \nonumber \\
& \quad + \left |  \prod_{i=1}^n \left (1 - \frac {t^2}{2}\V_h[\bar \bw_k(n,i)]\right ) - \prod_{i=1}^n  e^{-\V_h [\bar \bw_k(n,i)]t^2/2} \right |.
\label{eq:2mod}
\end{align}
Thus, we have to show that both moduli appearing in~(\ref{eq:2mod}) converge to zero. Consider the first. 
We know that $|\phi(t;n,i)| \le 1$, because the modulus of a characteristic function is never larger than unity. In addition, $\forall \epsilon>0$,
\begin{align}
&\V_h[\bar \bw_k(n,i)] = \E_h[\bar \bw_k^2(n,i)] \nonumber \\
& = \E_h\left [\bar \bw_k^2(n,i)  \cI_{| \bar \bw_k(n,i)|  \ge  \epsilon}\right ] + \E_h\left [\bar \bw_k^2(n,i)  \cI_{| \bar \bw_k(n,i)|  <  \epsilon} \right ] \nonumber \\
& \le \E_h\left [\bar \bw_k^2(n,i)  \cI_{| \bar \bw_k(n,i)|  \ge  \epsilon} \right ] + \epsilon^2 .
\label{eq:condvar1}
\end{align}
From~(\ref{eq:linde}) we see that 
\beq
\lim_{\eta_k\rightarrow 1} \lim_{n\rightarrow \infty}\max_{i=1,\dots,n} \E_h\left [\bar \bw_k^2(n,i)  \cI_{| \bar \bw_k(n,i)|  \ge  \epsilon} \right ] =0,
\label{eq:condvar2}
\eeq
which, used in~(\ref{eq:condvar1}), 
for the arbitrariness of $\epsilon$, gives
\beq
\lim_{\eta_k\rightarrow 1} \lim_{n\rightarrow \infty}  \max_{i=1,\dots,n} \V_h\left [\bar \bw_k(n,i)  \right ] = 0.
\label{eq:condvar3}
\eeq
As a consequence of~(\ref{eq:condvar3}), there exist a value of $n$ sufficiently large and a value of $\eta_k$ sufficiently close to one such that,
for all larger $n$ and $\eta_k$ closest to one, we have:
\beq
0 \le 1 - \frac {t^2}{2}\V_h[\bar \bw_k(n,i)] \le 1, \qquad i=1,\dots,n.
\label{eq:bo1}
\eeq 
Then,
we can apply the following inequality, valid for complex numbers $\xi_1,\dots,\xi_n$, $\zeta_1,\dots,\zeta_n$, satisfying $|\xi_i|\le 1$, $|\zeta_i|\le 1$, $i=1,\dots,n$:
\beq
\left |\prod_{i=1}^n \xi_i - \prod_{i=1}^n \zeta_i \right | \le \sum_{i=1}^n |\xi_i-\zeta_i|,
\label{eq:useb}
\eeq
see~\cite[Lemma 1, p.~358]{billingsley-book2}. This yields 
\begin{align}
& \left | \prod_{i=1}^n \phi(t;n,i) - \prod_{i=1}^n \left (1 - \frac {t^2}{2}\V_h[\bar \bw_k(n,i)]\right ) \right | \nonumber \\
& \le \sum_{i=1}^n \left | \phi(t;n,i)  - \left (1 - \frac {t^2}{2}\V_h[\bar \bw_k(n,i)]\right )
\right |, 
\end{align}
which converges to zero for $n\rightarrow \infty$ followed by $\eta_k\rightarrow 1$,
in view of~(\ref{eq:hopeok}).

Consider next the second modulus appearing in~(\ref{eq:2mod}). Fix an arbitrary $\epsilon>0$. We have: 
\begin{align}
&\left |  \prod_{i=1}^n \left (1 - \frac {t^2}{2}\V_h[\bar \bw_k(n,i)]\right ) - \prod_{i=1}^n  e^{-\V_h [\bar \bw_k(n,i)]t^2/2} \right | \label{eq:s0} \\
& \quad \le \sum_{i=1}^n \left | 1 - \frac {t^2}{2}\V_h[\bar \bw_k(n,i)] - e^{-\V_h [\bar \bw_k(n,i)]t^2/2} \right | \label{eq:s1} \\
& \quad \le \frac{t^4}{8}  \, \sum_{i=1}^n \V_h^2 [\bar \bw_k(n,i)]  \label{eq:s3} \\
& \quad \le \frac{t^4}{8}  \, \epsilon \sum_{i=1}^n \V_h [\bar \bw_k(n,i)]   = \frac{t^4}{8}  \, \epsilon. \label{eq:s4} 
\end{align}
In the above,~(\ref{eq:s1}) follows by using again~(\ref{eq:bo1}) and~(\ref{eq:useb}); 
inequality~(\ref{eq:s3}) is a consequence of the bound $\left | 1-z -e^{-z}\right | \le z^2/2$, valid for all $z\ge0$;
for any $\epsilon>0$, the inequality in~(\ref{eq:s4}) follows by~(\ref{eq:condvar3}); the equality in~(\ref{eq:s4}) follows from $\V_h [\bar \by_k(n)]=\sum_{i=1}^n \V_h [\bar \bw_k(n,i)] =1$. Since $\epsilon$ is arbitrary, we see that~(\ref{eq:s0}) goes to zero when $n\rightarrow \infty$ followed by $\eta_k\rightarrow 1$.

So far, we have shown that the characteristic function $\E_h e^{jt \sum_{i=1}^n \bar \bw_k(n,i)}$ of the random variable $\bar \by_k(n)$ converges to the characteristic function $\E[e^{jt\,\bg}]= e^{-t^2/2}$ of a standard Gaussian random variable $\bg$. 
Continuity theorem~\cite[p.~349]{billingsley-book2}  states that $\bar \by_k(n)$ converges in distribution to $\bg$ if, and only if, the characteristic function of $\bar \by_k(n)$ converges to the characteristic function of $\bg$, for each~$t \in \Re$. This shows that the distribution of $\frac{\by_k(n)-m_{k,h}(n)}{s_{k,h}(n)}$ converges to the standard Gaussian distribution, when $n \rightarrow \infty$ followed by $\eta_k \rightarrow 1$, which concludes the proof.

\section{Lyapunov Condition}
\label{app:lyapo}

The proof of Theorem~2 developed in Appendix~\ref{app:theo2} is founded on the following result.

\vspace*{5pt}
\noindent \textbf{Lemma 1 (Lyapunov condition):} 
\emph{Suppose that there exists $\beta>0$ such that $\E_h[|\bx|^{2+\beta}] < \infty$. Then, we have
\beq
\lim_{\eta_k\rightarrow 1}\lim_{n\rightarrow \infty} \sum_{i=1}^n \E_h \left [ |\bar \bw_k(n,i)|^{2 + \beta}   \right ] 
=0.
\label{eq:lyapo2}
\eeq}\hfill$\square$

\vspace*{5pt}\noindent
\emph{Proof.} 
For $z,\nu \in \Re$ with $\nu\ge1$, the function $|z|^\nu$ is convex. This implies, for $z_1,z_2 \in \Re$, 
\beq
\left | z_1 + z_2  \right |^\nu  \le 2^{\nu-1} \left ( |z_1|^\nu + |z_2|^\nu\right ).
\label{eq:pbound}
\eeq
From~(\ref{eq:defw2}), using the notations~(\ref{eq:note1}) and~(\ref{eq:note2}), and exploiting~(\ref{eq:pbound}) with $\nu=2+\beta$, one gets
\begin{align}
& |\bar \bw_k(n,i)|^{2+\beta}   = \left . \frac{\eta_k^{(i-1)(2+\beta)}}{s_k^{2+\beta}(n)} \; \;  \right | \mu \,  a_k \,  \bx^\prime_k(n-i+1) \nonumber \\
& \hspace*{100pt} \left . + \sum_{\ell \neq k} a_{k\ell} \widetilde \bx^\prime_{\ell}(n-i+1)  \right |^{2+\beta} \nonumber \\
& \le \left . 2^{1+\beta}\frac{\eta_k^{(i-1)(2+\beta)}}{s_k^{2+\beta}(n)} \right \{  \mu^{2+\beta} a_k^{2+\beta} \, \left |  \bx^\prime_k(n-i+1) \right |^{2+\beta} 
\nonumber \\
& \hspace*{100pt} + \left . \left |  \sum_{\ell \neq k} a_{k\ell} \widetilde \bx^\prime_{\ell}(n-i+1)  \right |^{2+\beta} \right \} \nonumber \\
& \le 2^{1+\beta}\frac{\eta_k^{(i-1)(2+\beta)}}{s_k^{2+\beta}(n)}  \left \{  \mu^{2+\beta} a_k^{2+\beta} \, \left |  \bx^\prime_k(n-i+1) \right |^{2+\beta} \right .
\nonumber \\
&  \hspace*{100pt} + \left .  M^{2+\beta} (1-a_k)^{2+\beta} \right \} \hspace*{-2pt}, \label{eq:aa}
\end{align} 
where $M/2 \dfz \max(|\E_0 \bx|,|\E_1 \bx|)$. In~(\ref{eq:aa}) we have exploited $|\sum_{\ell \neq k} a_{k\ell} \widetilde \bx^\prime_{\ell}|$ $\le$ 
$\sum_{\ell \neq k} a_{k\ell} |\widetilde \bx^\prime_{\ell}|$,
and $|\widetilde \bx^\prime_{\ell}|$ $\le$ $|\widetilde \bx_{\ell}| + |p \E_1 \bx+ (1-p) \E_0 \bx |$, with $p$ equal to 
$p_d$ under $\cH_1$, and equal to $p_f$ under $\cH_0$. This gives
$|\widetilde \bx^\prime_{\ell}|$ $\le$  $ |\widetilde \bx_{\ell}|+\max(|\E_0 \bx|,|\E_1 \bx|)$ $\le$ $M$, and therefore 
 $|\sum_{\ell \neq k} a_{k\ell} \widetilde \bx^\prime_{\ell}|$ $\le$  $\sum_{\ell \neq k} a_{k\ell} | \widetilde \bx^\prime_{\ell}|$
$\le M(1-a_k)$.
From inequality~(\ref{eq:aa}) it follows
\begin{align} 
&\sum_{i=1}^n \E_h  |\bar \bw_k(n,i)|^{2+\beta} \le  \sum_{i=1}^n 2^{1+\beta}\frac{\eta_k^{(i-1)(2+\beta)}}{s_k^{2+\beta}(n)} \nonumber \\
& \qquad \times \left \{  \mu^{2+\beta} a_k^{2+\beta} \, \E_h [\left |  \bx^\prime \right |^{2+\beta}] 
+ M^{2+\beta} (1-a_k)^{2+\beta} \right \} \nonumber \\
&= \frac{2^{1+\beta}}{s_k^{2+\beta}(n)}  \frac{1-\eta_k^{n(2+\beta)}}{1-\eta_k^{2+\beta}} \left \{  \mu^{2+\beta} a_k^{2+\beta} \, \E_h [\left |  \bx^\prime \right |^{2+\beta}] 
\right .  \nonumber \\
&  \hspace*{110pt} \left . + M^{2+\beta} (1-a_k)^{2+\beta} \right \}. \label{eq:sbound}
\end{align} 

Next, from~(\ref{eq:dfzs}):
\begin{align}
s_k^{2+\beta}(n)&=\left (\mu^2 a_k^2 \V_h \bx +  \V_h \widetilde \bx \, \sum_{\ell \neq k} a_{k\ell}^2\right )^{1+\frac \beta 2} \hspace*{-7pt}
\left (\frac{1-\eta_k^{2n}}{1-\eta_k^{2}} \right )^{1+\frac \beta 2}.
\label{eq:sagain}
\end{align} 
Applying Cauchy-Schwarz inequality  $(\sum_\ell {\xi_\ell \zeta_\ell})^2 \le \sum_\ell \xi_\ell^2 \sum_\ell \zeta_\ell^2$ to the real sequences 
$\xi_\ell=a_{k\ell}$ and $\zeta_\ell=1$, gives $\sum_{\ell\neq k} a_{k\ell}^2 \ge \frac{1}{S-1} (\sum_{\ell\neq k} a_{k\ell})^2=\frac{(1-a_k)^2}{S-1}$.
Using this latter inequality in~(\ref{eq:sagain}):
\begin{align}
s_k^{2+\beta}(n) \ge \left (\mu^2 a_k^2 \V_h \bx +  \V_h \widetilde \bx \, \frac{(1-a_k)^2}{S-1}\right )^{\hspace*{-4pt}1+\frac \beta 2} \hspace*{-5pt} \left (\frac{1-\eta_k^{2n}}{1-\eta_k^{2}} \right )^{\hspace*{-4pt}1+\frac \beta 2} \hspace*{-15pt},
\end{align} 
which, inserted in~(\ref{eq:sbound}), finally gives
\begin{align} 
\sum_{i=1}^n \E_h  |\bar \bw_k(n,i)|^{2+\beta} \le 2^{1+\beta} \Psi(\mu,a_k) \, \Gamma(\eta_k,n) ,
\label{eq:fingiv}
\end{align} 
where
\begin{align} 
&  \Psi(\mu,a_k) \dfz   \frac{  \mu^{2+\beta} a_k^{2+\beta} \, \E_h [\left |  \bx^\prime \right |^{2+\beta}] 
+ M^{2+\beta} (1-a_k)^{2+\beta} } 
{\left (\mu^2 a_k^2 \V_h \bx +  \V_h \widetilde \bx \, \frac{(1-a_k)^2}{S-1}\right )^{1+\frac \beta 2}}, \label{eq:finf} 
\end{align}
and
\begin{align}
& \Gamma(\eta_k,n)\dfz \frac{1-\eta_k^{n(2+\beta)}}{\left (1-\eta_k^{2n} \right )^{1+\frac \beta 2}}  \frac{\left (1-\eta_k^{2}\right)^{1+\frac \beta 2}}{ 1-\eta_k^{2+\beta}} . \label{eq:defg}
\end{align}

The function $\Psi(\mu,a_k)$ in~(\ref{eq:finf}) can be upper bounded as follows:
\begin{align} 
 \Psi(\mu,a_k) &= \frac{  \mu^{2+\beta} a_k^{2+\beta} \, \E_h [\left |  \bx^\prime \right |^{2+\beta}]} 
{\left (\mu^2 a_k^2 \V_h \bx +  \V_h \widetilde \bx \, \frac{(1-a_k)^2}{S-1}\right )^{1+\frac \beta 2}}  \nonumber \\
& \hspace*{50pt}  +\frac{  M^{2+\beta} (1-a_k)^{2+\beta} } 
{\left (\mu^2 a_k^2 \V_h \bx +  \V_h \widetilde \bx \, \frac{(1-a_k)^2}{S-1}\right )^{1+\frac \beta 2}}  \nonumber \\
&  \le \frac{  \mu^{2+\beta} a_k^{2+\beta} \, \E_h [\left |  \bx^\prime \right |^{2+\beta}]} 
{\left (\mu^2 a_k^2 \V_h \bx \right )^{1+\frac \beta 2}}  + \frac{  M^{2+\beta} (1-a_k)^{2+\beta} } 
{\left ( \V_h \widetilde \bx \, \frac{(1-a_k)^2}{S-1}\right )^{1+\frac \beta 2}} \nonumber \\
& = \frac{ \E_h [(\left |  \bx^\prime \right |^2)^{1+\frac \beta 2}]} 
{\left (\E_h [|\bx^\prime|^2] \right )^{1+\frac \beta 2}}  + \left (\frac{  M^2 (S-1)}  
{\V_h \widetilde \bx } \right )^{1+\frac \beta 2} \hspace*{-15pt},
\end{align} 
yielding $0 \le \Psi(\mu,a_k) \le \Psi_{\max}$, where
\beq 
\Psi_{\max} \dfz  \E_h \left [ \left ( \frac{|  \bx^\prime |^2}{\E_h [|\bx^\prime|^2]} \right )^{1+\frac \beta 2}\right ]
+ \left (\frac{  M^2 (S-1)}  
{\V_h \widetilde \bx } \right )^{1+\frac \beta 2} \hspace*{-15pt}. \label{eq:fmax}
\eeq

Now, it is easily seen that $\E_h[ |\bx^\prime|^2] >0$ and $\V_h \widetilde \bx>0$. The former is because $\E_h[ |\bx^\prime|^2] =0$ would imply $\bx=\E_h \bx$ with probability one, and we have instead assumed that $\bx$ has a density. The latter follows 
by the assumptions $\E_1 \bx \neq \E_0 \bx$ and $p_d,p_f \neq 0,1$.
In addition, the condition $\E_h [|\bx|^{2+\beta}] < \infty$ ensures
the finiteness of $\E_h [\left |  \bx^\prime \right |^{2+\beta}]$~\cite{FellerBookV2}. 
Therefore, $\Psi_{\max}$ in~(\ref{eq:fmax}) is finite: $\Psi_{\max} < \infty$.

As to $\Gamma(\eta_k,n)$ in~(\ref{eq:defg}), simple algebra shows that 
$0\le \Gamma(\eta_k,n) \le 1$, yielding,
from~(\ref{eq:fingiv}),
\beq
\lim_{n\rightarrow \infty}\sum_{i=1}^n \E_h  |\bar \bw_k(n,i)|^{2+\beta} \le  2^{1+\beta} \Psi_{\max}.
\label{eq:dp}
\eeq
Thus, the left-hand side of~(\ref{eq:dp}) is a bounded series of nonnegative terms, and therefore it converges to a finite limit. 
Now, using $\Psi(\mu,a_k) \le \Psi_{\max}$ in~(\ref{eq:fingiv}), we see that
\beq
\sum_{i=1}^n \E_h  |\bar \bw_k(n,i)|^{2+\beta} \le 2^{1+\beta} \Psi_{\max}\, \Gamma(\eta_k,n) .
\eeq
Therefore, from
\beq
\lim_{\eta_k\rightarrow 1}\lim_{n\rightarrow \infty}\Gamma(\eta_k,n) = \lim_{\eta_k\rightarrow 1}\frac{\left (1-\eta_k^{2}\right)^{1+\frac \beta 2}}{ 1-\eta_k^{2+\beta}}=0,
\label{eq:sit}
\eeq
we get~(\ref{eq:lyapo2}):
\begin{align} 
\lim_{\eta_k \rightarrow 1} \lim_{n \rightarrow \infty}   \sum_{i=1}^n \E_h  |\bar \bw_k(n,i)|^{2+\beta}  = 0.
\label{eq:lya}
\end{align}

It is worth noting that if one replaces the limit $\eta_k \rightarrow 1$ with the limit $\mu\rightarrow 0$, Lyapunov condition~(\ref{eq:lya})
does not hold anymore. This can be seen from~(\ref{eq:sit}) by observing that
\beq
\lim_{\mu\rightarrow 0}\lim_{n\rightarrow \infty}\Gamma(\eta_k,n)=\frac{(1-a_k^2)^{1+\frac\beta 2}}{1-a_k^{2+\beta}},
\eeq
which does not go to zero, unless the further condition $a_k\rightarrow 1$
is enforced.

\section{Choice of $\delta$ and $\bar m$ for the Exponential Example}
\label{app:delta}

Let us start by the second condition in~(\ref{eq:maxnew}): for some $\epsilon^{\prime}>0$,
\begin{empheq}[left=\empheqlbrace]{align}
&F_{\bu,h}\left(u -  \frac{2 \pi a_k \mu}{\delta}  \right ) \le \frac{\epsilon^{\prime}}{2}, \label{eq:firstc} \\
&1- F_{\bu,h}\left(u + \frac{2 \pi a_k \mu}{\delta}  \right ) \le \frac{\epsilon^{\prime}}{2}. \label{eq:secondc}
\end{empheq}

Since, with probability one, $\bx_k(\infty)\ge -\log \lambda_e$, from~(\ref{eq:winf})-(\ref{eq:uinf}) we see that 
$\bu_k(\infty)\ge - \frac{a_k \mu}{1-\eta_k}\log \lambda_e$. Then,
a sufficient condition ensuring~(\ref{eq:firstc}) is $u -  2 \pi a_k \mu/\delta  \le -\frac{a_k \mu}{1-\eta_k}\log \lambda_e$. Assuming $u \ge -\frac{a_k \mu}{1-\eta_k}\log \lambda_e$ [otherwise~(\ref{eq:firstc}) is automatically verified for all $\delta>0$], this yields:
\beq
\delta \le \frac{2 \pi }{\frac{u}{a_k \mu} + \frac{\log \lambda_e}{1-\eta_k}}.
\label{eq:du1}
\eeq

As to~(\ref{eq:secondc}), exploiting Chebyshev inequality~\cite{papoulis}, we have
\begin{align}
&\,\,\, 1- F_{\bu,h}\left(u + \frac{2 \pi a_k \mu}{\delta}  \right ) = \P_h \left( \bu_k(\infty) > u + \frac{2 \pi a_k \mu}{\delta}  \right ) \nonumber \\
& = \P_h \left(  \bu_k(\infty) - \E_h[\bu_k(\infty)]  > u -  \E_h[\bu_k(\infty)]  + \frac{2 \pi a_k \mu}{\delta}  \right ) \nonumber \\
&  \le \P_h \left( \left | \bu_k(\infty) - \E_h[\bu_k(\infty)]  \right |> u - \E_h[\bu_k(\infty)] + \frac{2 \pi a_k \mu}{\delta}   \right ) \nonumber \\
&  \le \frac{\V_h[\bu_k(\infty)]}{\left ( u - \E_h[\bu_k(\infty)] + 2 \pi a_k \mu/\delta \right )^2}, \label{eq:c2}
\end{align} 
where we have assumed 
$u - \E_h[\bu_k(\infty)] + \frac{2 \pi a_k \mu}{\delta} > 0$. 
Now, the last line of~(\ref{eq:c2}) is not larger than $\epsilon^{\prime}/2$ if
\beq
\delta \le \frac{2 \pi a_k \mu}{\sqrt{\frac{2 \V_h[\bu_k(\infty)]}{\epsilon^{\prime}}}+\E_h[\bu_k(\infty)]-u},
\label{eq:du2}
\eeq
assuming that right-hand side of~(\ref{eq:du2}) is positive. 

Admittedly, bounds~(\ref{eq:du1}) and~(\ref{eq:du2}) are loose, but sufficient for our purposes.
The criterion used in the numerical experiments is to fix $\epsilon^{\prime}$, and then: if $u \ge -\frac{a_k \mu}{1-\eta_k}\log \lambda_e$ we choose $\delta$ as the minimum between the right-hand sides of~(\ref{eq:du1}) and~(\ref{eq:du2}), otherwise we choose $\delta$ equal to the right-hand side of~(\ref{eq:du2}). 
In our experiments $\epsilon^\prime= 2 \, 10^{-5}$. For this value we have verified numerically that the error on $F_{\bu,h}(u)$ due to the series truncation [see first condition in~(\ref{eq:maxnew})]  is negligible.
Finally, the value of $\bar m$ is selected using the empirical criterion $\eta^{\bar m} \le \epsilon^{\prime \prime}$, for some sufficiently small $\epsilon^{\prime \prime}$. Namely, we set $\bar m =\lceil \frac{\log \epsilon^{\prime \prime}}{\log \eta_k} \rceil$, and in the numerical experiments we use 
$\epsilon^{\prime \prime}=\epsilon^{\prime} = 2 \, 10^{-5}$.

\end{appendices}



\end{document}